\documentclass[letterpaper,11pt]{article}
\usepackage{fullpage}

\usepackage[parfill]{parskip}
\usepackage[letterpaper, left=1in, right=1in, top=1in, bottom=1in]{geometry}

\usepackage{natbib}
 \bibpunct[, ]{(}{)}{,}{a}{}{,}%
\usepackage{pgfplots}
\usepackage{tikz}
\usepackage[caption=false]{subfig}
\usetikzlibrary{topaths,calc}

\usepackage{amsmath,amsthm,amssymb}

\usepackage{xfrac}

\usepackage{comment}
\usepackage{bbm}
\usepackage{hhline}
\usepackage{mathtools}
\usepackage{ifthen}

\usepackage{graphicx,color}
\usepackage{fancybox}

\usepackage{array,float}
\usepackage{url}
\usepackage{mathrsfs}

\definecolor{cornellred}{rgb}{0.7, 0.11, 0.11}
\definecolor{maroon}{rgb}{0.52, 0, 0}
\definecolor{dgreen}{rgb}{0.0, 0.5, 0.0}
\definecolor{ballblue}{rgb}{0.13, 0.67, 0.8}
\definecolor{royalblue(web)}{rgb}{0.25, 0.41, 0.88}
\definecolor{bleudefrance}{rgb}{0.19, 0.55, 0.91}
\definecolor{royalazure}{rgb}{0.0, 0.22, 0.66}

\usepackage{enumitem}
\usepackage{multirow}
\usepackage{changepage}

\usepackage{accents}
\usepackage{todonotes}

\usetikzlibrary{arrows, decorations.markings}

\definecolor{bostonuniversityred}{rgb}{0.8, 0.0, 0.0}
\usepackage{hyperref}
\hypersetup{
	colorlinks = true,
	linkcolor=blue!70!black,
	citecolor=blue!70!black,
	linkbordercolor = {white}
}
\usepackage{cleveref}

\usepackage{thmtools}
\theoremstyle{plain}
\newtheorem{theorem}{Theorem}[section]
\newtheorem{lemma}[theorem]{Lemma}
\newtheorem{claim}[theorem]{Claim}

\newtheorem{proposition}[theorem]{Proposition}
\newtheorem{corollary}[theorem]{Corollary} 

\usepackage{thm-restate}

\theoremstyle{plain}
\newtheorem{definition}{Definition}[section] 
\newtheorem{example}[definition]{Example}

\theoremstyle{plain}

\newcommand{\xhdr}[1]{\vspace{8pt}\noindent{\bf {#1.}}}

\newcommand{\omt}[1]{}

\newcommand{\squishlist}{
   \begin{list}{{{\small{$\bullet$}}}}
    { \setlength{\itemsep}{3pt}      \setlength{\parsep}{1pt}
      \setlength{\topsep}{1pt}       \setlength{\partopsep}{0pt}
     \setlength{\leftmargin}{1em} \setlength{\labelwidth}{1em}
      \setlength{\labelsep}{0.5em} } }
\newcommand{\squishend}{  \end{list}  }

\newcommand{\squishlistTmp}{
   \begin{list}{{{\small{$\bullet$}}}}
    { \setlength{\itemsep}{2pt}      \setlength{\parsep}{0.5pt}
      \setlength{\topsep}{0.5pt}       \setlength{\partopsep}{0pt}
     \setlength{\leftmargin}{0.5em} \setlength{\labelwidth}{0.5em}
      \setlength{\labelsep}{0.5em} } }
\newcommand{\squishendTmp}{  \end{list}  }

\newcounter{relctr} 
\everydisplay\expandafter{\the\everydisplay\setcounter{relctr}{0}} 
\newcommand\labelrel[2]{%
  \begingroup
    \refstepcounter{relctr}%
    \stackrel{\textnormal{(\alph{relctr})}}{\mathstrut{#1}}%
    \originallabel{#2}%
  \endgroup
}
\AtBeginDocument{\let\originallabel\label} 

\newcommand{\argmax}{\mathop{\mathrm{arg\,max}}}

\newcommand{\N}{\mathbb{N}}

\newcommand{\E}{\mathbb{E}}

\usepackage{relsize}

\newcommand*{\R}{\mathbb{R}}

\DeclareMathOperator{\OPT}{\texttt{OPT}}

\newcommand{\messagespace}{\mathcal{M}}

\newcommand{\messageprob}{\Phi}

\newcommand{\signal}{q}

\newcommand{\cc}[1]{\ensuremath{\mathsf{#1}}} 

\newcommand{\agentU}{u^A}  
\newcommand{\senderU}{u^S}  

\newcommand{\senderURealize}{U^S}

\newcommand{\cost}{cost}  

\newcommand{\blackwellOrder}{\succeq}

\newcommand{\infoStrategy}{strategy}
\newcommand{\infoStrategies}{strategies}

\newcommand{\essienFull}{essentially full information strategy}

\newcommand{\fullInforStrategy}{full information strategy}
\newcommand{\noInforStrategy}{no information strategy}

\newcommand{\priorCDF}{H}

\newcommand{\NI}{\cc{NI}}
\newcommand{\openIdx}{\mathbb{I}}
\newcommand{\selectIdx}{\mathbb{A}}

\newcommand{\condition}{\,\mid\,}

\newcommand{\prob}[2][]{\text{Pr}\ifthenelse{\not\equal{}{#1}}{_{#1}}{}\!\left[{\def\givenn{\middle|}#2}\right]}
\newcommand{\expect}[2][]{\mathbb{E}\ifthenelse{\not\equal{}{#1}}{_{#1}}{}\!\left[{\def\givenn{\middle|}#2}\right]}

\newcommand{\tparen}{\big}
\newcommand{\tprob}[2][]{\text{Pr}\ifthenelse{\not\equal{}{#1}}{_{#1}}{}\tparen[{\def\given{\tparen|}#2}\tparen]}
\newcommand{\texpect}[2][]{\mathbb{E}\ifthenelse{\not\equal{}{#1}}{_{#1}}{}\tparen[{\def\given{\tparen|}#2}\tparen]}

\newcommand{\sprob}[2][]{\text{Pr}\ifthenelse{\not\equal{}{#1}}{_{#1}}{}[#2]}
\newcommand{\sexpect}[2][]{\mathbb{E}\ifthenelse{\not\equal{}{#1}}{_{#1}}{}[#2]}

\newcommand{\indicator}[1]{{\mathbbm{1}\left\{ #1 \right\}}}

\newcommand{\supp}[1]{{\texttt{supp}\left[#1 \right]}}

\newif\ifOR
\ORtrue
\newif\ifTE
\TEfalse

\ifodd 1
\newcommand{\wt}[1]{{\color{red}(WT: #1)}}
\newcommand{\ding}[1]{{\color{red}(DING: #1)}}
\newcommand{\cj}[1]{{\color{blue}(CJ: #1)}}

\newcommand{\hfr}[1]{{\color{black}  #1}}
\else
\newcommand{\wt}[1]{}
\newcommand{\ding}[1]{}
\newcommand{\cj}[1]{}
\fi
\newcommand{\ignore}[1]{}

\newcommand{\wtr}[1]{{\color{black}  #1}}
\newcommand{\wtrTE}[1]{{\color{black}  #1}}

\title{Competitive Information Design for Pandora's Box}

\author{
Bolin Ding\thanks{Alibaba Group, 
\texttt{bolin.ding@alibaba-inc.com}} \and 
Yiding Feng\thanks{Hong Kong University of Science and Technology, 
\texttt{ydfeng@ust.hk}} \and 
Chien-Ju Ho\thanks{Washington University in St. Louis, \texttt{chienju.ho@wustl.edu}} \and
Wei Tang\thanks{Chinese University of Hong Kong, \texttt{weitang@cuhk.edu.hk}} \and
Haifeng Xu\thanks{University of Chicago, 
\texttt{haifengxu@uchicago.edu}}  
}

\begin{document}
\maketitle

\begin{abstract}

We study a natural competitive-information-design strategic variant for the celebrated Pandora’s Box problem \citep{wei-79}, 
where each box is associated with  
a strategic information sender who can design 
what information about the box's prize value to be revealed to the agent when the agent inspects the box. 
This variant with strategic boxes is motivated by 
 a wide range of real-world economic applications for Pandora's Box.  
\wtrTE{Our contributions  are three-fold: 
\wtrTE{(1) given the boxes' information policies, we characterize the agent's optimal search and stopping strategy;}
(2) we fully characterize the pure symmetric equilibrium for the game of boxes' competitive information revelation 
\wtr{in a symmetric environment};
and (3) we reveal various insights regarding information competition and the resultant agent payoff at equilibrium, 
and additionally, we study informational properties of Pandora's Box by establishing an intrinsic connection between \emph{informativeness} of any box's value distribution and the \emph{utility order} of the search agent.}
\footnote{\footnotesize{\emph{An earlier conference version of this work has appeared in the proceeding of the ACM-SIAM Symposium on Discrete Algorithm (SODA'23)~\citep{BYCWH-23}.}}}

\end{abstract}

\newpage


\section{Introduction}

\newcommand{\condiOne}{\textbf{Maximum reservation value}}
\newcommand{\condiTwo}{\textbf{$G$'s shape below reservation value}}
\newcommand{\condiThree}{\textbf{No deviation incentive}}
\newcommand{\deviationRV}{\sigma^*}


The Pandora's Box problem, 
as formalized in the seminal work of  \citet{wei-79}, 
is a foundational framework for 
studying 
 how the cost of acquiring information 
affects the adaptive decisions about what   information to acquire ---  
the  obtained information from the past  
will affect whether additional information is needed,  and if so which information to  acquire  next. 
Specifically,  the Pandora's Box problem is described as follows. 
An agent is presented with $n$ boxes; each contains an unknown random prize. 
The value of the prize inside each box
is independently 
sampled from its distribution. While the agent knows each box's prize distribution, he does not know its realized value. Nevertheless, the agent can open any box (in any order) to learn its \emph{realized} prize value but suffers an associated opportunity cost for opening the box. The agent can stop at any time and claim one prize from some opened box, upon which the game terminates.   
The agent's goal is to maximize the expected prize value 
minus the total box-opening costs.
This basic model finds applications in numerous economic applications and thus,  unsurprisingly, 
has been extensively studied in  the economics, operations research, 
and computer science literature.  
For example, 
in house hunting,
a home buyer  incurs  cost to search for information 
about each potential house (e.g., attending its open house) and, at some point,   
 decide  to purchase one of the searched house and terminate the procedure. 
Similarly, many online customers   spend time 
on  free trials to obtain information about different digital services and, at some point,  decide to subscribe to some tried service.  

A surprisingly  simple and elegant policy
provided by \citet{wei-79} has been shown to be optimal for the Pandora's Box problem, despite its seemingly complex sequential decision process.  
Specifically, \citet{wei-79} defines certain {\em reservation value}  for each box, which is determined by  both the box's prize distribution and opening cost. The optimal policy simply sorts boxes in decreasing order of their reservation values, and then open boxes in this order until the thus-far maximum realized prize value 
exceeds the next box's reservation value. The agent then terminates the search by selecting that maximum realized prize.

An important assumption of the classic Pandora's Box problem ---  which is the one we intend to relax in this work --- is that  each box is an inanimate object and, once opened, will fully disclose its realized prize to the agent. Yet this may not be the case in many real-world applications where boxes often correspond to real strategic agents who may have incentives to selectively disclose information for their own interest \citep{M-21, BL-18, AR-06}. This is usually the case when information is not controlled by nature but by humans or algorithms. The following are two of many such examples.

\begin{example}[Open Houses in Housing Markets]
During open houses, many house sellers typically would design events to highlight their house qualities and these event schedules will be sent to potential buyers. This corresponds to the boxes' design and 
commitment to an information disclosure policy. 
Informed with these policies (i.e., learning what he expects to see), 
a buyer will decide which open houses to visit in what sequence, and during this process the buyer may make a purchase decision (i.e., stop searching). 
In this example,
it is costly for a buyer to obtain the information from any box due to the time spent to travel and visit.
Moreover, the seller usually selectively discloses information in order to maximize the chance of sale.
Built upon Weitzman's elegant solution to the classic Pandora's box for the buyer's search, our work studies the house sellers' competitive information design problem and how sellers'  revealed information affects the agent's total utility.
\end{example}

\begin{example}[Free Trials of Digital Services]
Consider online services like YouTube Music, Spotify, and Amazon Music. To attract users for subscription, these services often offer   free trials (e.g., an one-month free trial with access to a limited set of functionalities of the service) before the user picks one service to subscribe. 
These free trials, including the functionalities included in this period, 
can be seen as a committed information revelation policy designed by the service provider.  
The user needs to pay search costs (i.e., time spent to explore) 
to obtain the information. 
Moreover, these information policies are usually not full-information revealing due to limited trial periods or limited functionality access. 
In contrast to the fully observable prize value in classic Pandora's Box, the user here can only form an updated belief about the service quality before choosing a subscription. 
\end{example}
Motivated by real-world applications like the above, this paper studies a natural information design variant of the celebrated Pandora's Box problem by viewing each box as an economic agent with its own actions and incentives. We assume that, before the agent opens any box, each box commits to an  information revelation policy --- a.k.a., a signaling mechanism which stochastically maps the underlying  prize to a random \emph{signal} --- to selectively disclose  information 
about the prize. Afterwards,  the agent   engages 
in a costly search across boxes, i.e., solving a standard Pandora's Box problem, in order to collect the most-rewarding prize in expectation. Notably, after opening any box, the agent now is only able to observe a realized signal that carries partial information about the underlying prize value, but cannot directly observe the prize value.

We study a model where there are 
$n$ boxes, competing with each other 
for being selected by the agent.
The agent is assumed to initially hold the common prior belief $H_i$ about the prize distribution of each box $i\in[n]$.
We assume boxes are decentralized (e.g., corresponding to different product sellers). Each box can flexibly choose any signaling mechanism to strategically reveal information about his own prize. This gives rise to a natural competitive information design problem in the Pandora's Box with many senders, e.g., the boxes.
\wtrTE{The main focus of this paper is how the boxes design the signal mechanisms and how these mechanisms subsequently shape the agent's search and stopping behavior, ultimately affecting the box’s own payoff.
To this end, we focus on subgame-perfect Nash equilibrium solution concept of this game with multiple leaders (i.e., the boxes) and a single follower (the agent).}


\subsection{Our Contribution}
Our contributions  are three-fold: 
\wtrTE{(1) given the boxes' information policies, we characterize the agent's optimal search and stopping strategy;}
(2) we fully characterize the pure symmetric equilibrium for the boxes' competitive information revelation 
\wtr{in a symmetric environment};
and (3) we reveal various insights regarding information competition and the resultant agent payoff at equilibrium,
and additionally, we study informational properties of Pandora's Box by establishing an intrinsic connection between \emph{informativeness} of any box's value distribution and the \emph{utility order} of the search agent.
%

\wtrTE{\xhdr{Agent's optimal strategy}
When the boxes do not strategically reveal information about the prize value, it is well-established that the agent’s optimal strategy follows the reservation-value-based approach introduced by \cite{wei-79}.
In our setting, some signaling mechanisms may be more or less informative than others, and the agent can only observe a noisy signal about the underlying prize value when inspecting the boxes.
Thus, it is unclear in what order the agent should inspect the boxes or when to stop, based on the observed signals.
Perhaps surprisingly, we show that when the agent is risk-neutral, even though she faces uncertainty about the prize value during box inspections, her optimal strategy still follows a structure similar to \cite{wei-79}'s approach. 
In particular, each box's signaling mechanism can be viewed as a distribution of posterior means of the prize value. Under the optimal strategy, the game proceeds as if each box reveals the prize value according to its posterior mean distribution, and the agent searches using Weitzman's strategy applied to these posterior mean distributions.

A nice consequence of this characterization of the agent's optimal strategy is that we can, without loss of generality, reformulate each box's signaling mechanism as a distribution of posterior means.
At first glance, this reformulation seems a bit surprising, given the complex interactions between the signaling mechanisms and the agent's search strategy. However, we show that under agent's optimal strategy, the agent determines the search order solely based on the posterior mean distributions and decides when to stop based on the realized posterior means. This search behavior justifies the reformulation of the boxes' signaling mechanisms as posterior mean distributions, without any loss of generality.} 

\xhdr{Equilibrium characterizations}
Our next main result is to identify
a necessary and sufficient condition
for the existence of a pure symmetric Nash equilibrium 
\wtr{when all boxes are all ex-ante symmetry.}
Moreover, if a pure symmetric equilibrium exists, our result provides a straightforward, and also computationally tractable, way to 
identify the equilibrium strategy. 
Specifically, we show that a pure symmetric equilibrium strategy $G$, if exists,
 must be fully characterized by the following three conditions:
\begin{itemize}
    \item[$(i)$]
    \condiOne: strategy $G$ must have maximum reservation value.
    \item[$(ii)$]
    \condiTwo:
    function $G^{n-1}$ is convex over its support, 
    and 
    linear whenever the strategy $G$ does not equal to the prior $H$,
    \wtr{where $H \equiv H_i, \forall i\in[n]$}.
    \item[$(iii)$]
    \condiThree:\
    there exists a reservation value $\deviationRV$ such that 
    deviating to a \infoStrategy\ that has 
    this reservation value $\deviationRV$ is not profitable.
\end{itemize}
Note that in our Pandora's Box problem, each box can design 
his information strategy to endogenously affect the agent's 
inspection order of boxes. 
The above condition $(i)$ ensures that, in equilibrium,
each box prefers to be inspected by the agent earlier rather than later
(recall that once the boxes' information strategies are fixed, 
the agent's optimal inspection strategy 
is to open the box in an decreasing order 
of their reservation values). 
The condition $(ii)$ then specifies the behavior 
that is below the corresponding reservation value 
of the equilibrium strategy.
We prove that the first two conditions above
can already uniquely pin down a
strategy as an equilibrium candidate. 
Core to our characterization is the third condition 
which verifies whether this strategy candidate 
is indeed an equilibrium or not.
The verification in condition $(iii)$, including the 
reservation value $\deviationRV$,
has a closed form and can be easily computed given the 
structure of the identified strategy $G$ from   conditions $(i)$ and $(ii)$.

We highlight two predominant challenges in deriving our main result on equilibrium characterizations, followed by our approaches 
to tackle these challenges.  
First, to see whether a strategy profile $(G, \ldots, G)$ 
is an equilibrium, we need to argue that no box has a
profitable deviation under this strategy profile. 
A box's best response problem can be formulated
as a linear program, after fixing all other boxes' \infoStrategies\ to be $G$. 
Prior works \citep{AK-20,HKB-19} have investigated a special case of our setting
where there is no \cost\ and the agent observes all realized prizes. 
They
have utilized this linear program approach to demonstrate that
the box's best response \infoStrategy\ is indeed $G$ itself
if $G$ is a certain equilibrium strategy candidate.
Note that in their setting,
no matter what the response \infoStrategy\ is,
the box's expected payoff
when realizing prize with value $x\in[0, 1]$ has a succinct
and well-structured form: 
$G(x)^{n-1}$. 
However, in our setting, different \infoStrategies\
have different reservation values, which impact
the order of the agent inspecting the box, and thus 
making the box's payoff function 
different and more complex. 
Consequently, there is no single linear program 
that can characterize a box's best response problem. 
Instead, for each possible reservation value $\sigma$, 
we consider a corresponding linear program which 
characterizes the best response \infoStrategy\ 
subject to a constraint that it 
has the same reservation value $\sigma$ 
(requiring a \infoStrategy\ to have a 
reservation value $\sigma$ can be formulated as 
a linear constraint).
We then prove that the optimal objective value of the
linear program, as a function of the given reservation value $\sigma$, is  a single-peaked function with the peak
achieved at some $\deviationRV$.

Second, for any reservation value $\sigma$, solving 
its corresponding linear program 
(i.e., the program to solve a box's best response problem) 
is  highly non-trivial. 
Let $F$ denote the response strategy 
used by the box and all other boxes use the strategy $G$.
There are two major constraints in this program:
one constraint accounts for the feasibility of the
\infoStrategy\ $F$, i.e., 
$H$ is an MPS of $F$; and
the other accounts for 
the reservation value constraint as it requires that the 
reservation value of strategy $F$ 
equals to $\sigma$.
\citet{DM-19} developed  an optimality verification technique based on strong duality for the special case with only the first constraint   (later employed by  \citealp{HKB-19}). Unfortunately, this technique does not directly apply to our more general case in presence of the second constraint as well. To overcome this barrier, we generalize the approach in \citep{DM-19}  to account for the additional constraint and 
characterize
corresponding optimal dual solution (of a new format).
This then allows us to verify the optimality of certain desired information structure based on the complementary slackness.


\xhdr{Informational properties of Pandora's Box 
and the agent's payoff} 
\wtrTE{Having established the agent's optimal strategy,}
we also show an intrinsic connection between informativeness of any box's value distribution and the utility order of the search agent. 
Formally, we prove that a distribution $H$ is more informative than $G$ in the Blackwell sense if and only if  in an arbitrary (not necessarily symmetric) Pandora'x Box's problem with $H$ as some box's value distribution, the agent's expected payoff weakly increases  when  this box's value distribution  switches  from $H$  to $G$.  This result complements a  fundamental result of  \cite{Blackwell-53}:  i.e., a distribution $H$ is more informative than a distribution $G$ if and only if $\expect[x\sim H]{u(x)} \geq \expect[x\sim G]{u(x)}$ for 
any convex function $u$. 
Since any convex 
function corresponds to a \emph{static} Bayesian decision making problem, Blackwell's result is viewed as a decision-theoretic foundation for informativeness of a distribution. Our result extends this  insight to a basic setup of \emph{sequential} decision making.
We remark that it is not obvious in hindsight that  more information from \emph{any} box would always benefit the agent. Recall that 
the agent's optimal inspection strategy depends on 
the order of reservation values of boxes' \infoStrategies. 
To prove the above result, we first show that the reservation value of a box always weakly increases
when the box's distribution becomes 
 more informative. Thus if a box with very bad expected prize value 
becomes more informative, this box's reservation value will  increase and thus it will be inspected early. However, it is not clear whether inspecting such a ``bad'' box earlier by lowering the priority of  other possibly better boxes will always benefit the agent since this may delay the agent's stop time and thus lead to increased cost. Our main result gives an affirmative answer. Our proof heavily hinges on    various properties of MPS in order to argue that the benefit of getting more information from any box can offset the possible harm of lowering the priority of other boxes.

A natural corollary of the above  result 
in our competitive information design environment
is that, 
when all boxes fully reveal the information about 
their prizes, the agent obtains the highest expected payoff.
Nevertheless, we strengthen this observation by showing that the agent can derive
the highest expected payoff as long as each box 
use a \infoStrategy\ which reveals full information
whenever the value of the prize is \emph{below} 
its reservation value.
We refer to this 
class of \infoStrategies\ as {\em \essienFull}. 
We  provide necessary and sufficient conditions on
when this strategy is the equilibrium strategy next. 

Next we describe additional insights conveyed by the above main result and discuss how the competition 
and the agent's \cost\  affect the boxes' equilibrium strategy.
Utilizing our conditions above, we can show that
\essienFull\ is the
equilibrium strategy if and only 
if function $H^{n-1}$ is convex in $[0, \sigma_H]$
where $\sigma_H$ is the reservation value of the distribution $H$.
Build upon this result, we are able to show that 
the \essienFull\ is more likely to become the equilibrium strategy when increasing the competition (i.e., increasing the number of boxes) 
or increasing the \cost.
The former is because, intuitively, increasing competition
``convexifies'' the shape of function $H^{n-1}$ and 
makes the condition more likely to be satisfied.  
The later is because the \cost\ affects  the reservation value $\sigma_H$ and thus the structure 
of (possible) equilibrium strategy 
$G$. 
First, we can see that
the \essienFull\ is the equilibrium strategy
under a larger \cost\
if it is already the equilibrium 
strategy under a smaller cost.
This is due to the monotonicity of reservation value $\sigma_H$
over the cost, i.e., a larger cost leads to a smaller $\sigma_H$. 
Second, as the \cost\ goes to $0$, 
the above characterized behavior of $G$ 
below its reservation value in condition $(ii)$
spans to the whole interval $[0, 1]$.\footnote{To ease exposition consider that
the value of prize is in $[0, 1]$.}
Third, the cost also plays a role in condition $(iii)$ 
as it determines the choice of reservation value $\deviationRV$.

\subsection{Related Work}
Our paper studies an information design variant 
of Pandora's Box.
The information design part follows the Bayesian persuasion setup by~\citet{KG-11}.
Their work has inspired an active line of research in information design games in various applications
(e.g., see the surveys by \citealp{Kamenica-19,BM-19} for economics literature and \citealp{D-17} for computer science literature).
Our work complements this line of research by exploring the competition in information design in the setup of Pandora's Box and discusses how the senders' signaling mechanisms shape the agent's searching behavior and ultimately affecting senders their own payoff.
Since there are multiple boxes designing the information strategies, our paper relates closely to the works in the multi-sender Bayesian Persuasion literature
\citep{GK-16,GK-17,GHHS-22}.
\wtrTE{In particular, the equilibrium analysis part of our work relates to the works \citep{BC-18,AK-19,AK-20,HKB-19} that also study a game with {\em ex ante} symmetric senders.  
Our work differs from theirs as they focus on a non-search setting where there is no inspection cost for the agent, and the agent can simply observe all realized values and then select a best one.}

\wtrTE{Perhaps the most closely related works are \cite{AW-23,HL-21,BL-18,HHS-20,HHS-22}, all of which also explore the (competitive) information design problem in search settings.
In particular, \citet{AW-23} examine a search setting similar to ours but focus on a simplified case where the prize value is binary, whereas we address the more involved case of a continuously distributed prize.
The continuous prize setting significantly complicates the analysis:
First, in the binary prize setting, for any agent's search strategy, it is straightforward to reformulate the boxes' signaling mechanisms as distributions of posterior means without loss of generality, which can greatly simplify subsequent analysis.
In contrast, this reformulation is not generally valid in the continuous prize setting for arbitrary search strategies.
Despite this challenge, we are able to establish such reformulation by a careful characterization of agent's optimal strategy.
Second, unlike the binary prize setting, our equilibrium analysis involves solving a best-response optimization problem with a second-order stochastic dominance constraint, which makes our analysis more involved. 
\cite{HL-21} study a search setting where the agent uses a random search strategy, while ours focus on optimal agent's search strategy.
\cite{BL-18} study a setting where sellers compete by designing experiments and buyers search sequentially. In contrast to our work, the sellers' experiments in their work are not publicly posted and cannot shape
the buyers' search.
In our setting, however, the agent's search behavior is directly influenced by the signaling mechanisms (or ``experiments'') of the boxes.
\citet{HHS-20,HHS-22,XHW-22} concern a setting that there exists a central planner that can collect 
all information from all boxes and then strategically reveal these information to the search agent, 
while in our setting,  each box itself is decentralized to be strategic and competes with each other for the final choice of the search agent. }

Our work on focusing the competition among
boxes relates to the literature about the market competitions.
For example, similar to our work, \citet{CDK-18} also
consider an oligopoly model in which consumers engage in sequential search for the best product 
based on partial product information (and prices), 
and the authors provide sufficient conditions that guarantee the  existence and uniqueness of market equilibrium. 
A notable difference to our work  is that the partial product information,  in our setting, is endogenously determined  by the boxes (a.k.a., sellers) themselves,
while in \cite{CDK-18}, the product information is exogenously fixed.
\citet{BCT-19} study a setting where 
a monopolistic information provider who can sell 
potentially informative signals to a collection of sellers that compete with one  another in a downstream market.

\ifTE
\else
\fi

We also mention recent 
technical developments on using the duality theory to 
characterize the optimal persuasion scheme 
in information design. In particular, 
\citet{DM-19,K-18} study the sender's problem on how to optimize 
the sender's (indirect) payoff as a function
of expected value (state) he induces, subject to the
feasible information strategy constraint.
Our work differs from theirs as we 
study the equilibrium in a strategic environment.
Moreover, though we can write the box's payoff as
a function of the expected prize value, this payoff
function further depends on the reservation value of 
the box's \infoStrategy\
(and other boxes' reservation values), and thus, their results does not apply directly. 
Instead, we extend their results to account for the additional reservation value constraint, and use the extended results to characterize the  optimal dual (primal) solution.

This paper is built on the seminar work of Pandora's Box introduced by~\citet{wei-79},  which, together with the prophet inequality, has been widely used to model the sequential search and stopping process under uncertainty in various domains (see, e.g., \citealp{OW-15,KWE-16,D-18,BK-19,CGTTZ-20,BFLL-20, FLL-23,CCES-24,CCFOT-24}).
Our work significantly differs from the previous works as we focus on
the boxes' behavior on strategically disclosing the prize 
information to the agent.

\section{A Model of Competitive Information Design for Pandora's Box}

In this section, we first revisit the formulation
of the classic Pandora's Box problem, and then formally introduce our setting as 
its natural  variant with competitive information design. 

\xhdr{The Pandora's Box problem}
In the Pandora's Box problem, a risk-neutral  
agent is presented with a set of $n$ boxes. 
Each box $i \in [n]$ contains a prize of value $x_i\in[0, 1]$.
The value $x_i$ is distributed according to a distribution
$G_i$, independent of the values of other boxes. 
For each box $i$, the agent does not know the value $x_i$ but knows the value distribution $G_i$. 
Moreover, the agent can pay a cost $c_i$ to inspect box $i$ and observe the value $x_i$.
The agent can choose to inspect any number of boxes in any order and take one 
of the values from the inspected boxes.
The goal of the agent is to maximize the value from the chosen box minus the total cost for inspecting boxes.

The agent's strategy $\pi$
is a rule that determines adaptively, 
at any time $t \ge 0$, 
whether to terminate the inspection and,
if not, which box to inspect next. 
The strategy also determines 
which box to select after the inspection ends.
Given a strategy $\pi$, let $\openIdx_i$ denote the indicator for whether box $i$ is inspected and 
$\selectIdx_i$ denote the indicator for whether box $i$ is chosen according to $\pi$.
The agent's goal is to choose a strategy 
$\pi$ which maximizes the following expected payoff
\begin{align}
    \label{agent_exp_payoff}
    \E\left[\sum_i \left[\selectIdx_i x_i - \openIdx_i c_i\right]\right]~.
\end{align}
Importantly, the agent can {\em only} claim one prize 
but must pay for all inspection costs.

\xhdr{Pandora's Box with competitive information design}
In this paper, we consider a natural competitive
information design variant
of the Pandora's Box problem, which is fundamentally a multi-leader (boxes) and single-follower (the agent) Stackelberg game.
Specifically, each box is associated with a strategic sender\footnote{In the following discussion, we interchangeably use ``box'' and ``sender''.}
who can design 
what information about the prize value 
the agent will see when she inspects the box. 
Similar to the classic problem, 
the agent does not know the values in boxes but holds some prior beliefs 
about the distribution of each value $x_i$.
However, different from the classic problem,
when the agent pays a cost to inspect box $i$,
she does not directly observe the value $x_i$. 
Instead, she observes some information signal, designed by the sender of box $i$, that is related to the prize $x_i$.
Following the literature in information design, 
this can be formalized as follows:
each sender $i$ is associated with an {\em ex ante} identical prize distribution $\priorCDF_i\in\Delta([0,1])$,\footnote{Our results can be readily generalized to an arbitrary interval $[a, b]$. To simplify the presentation, in this paper, we restrict our attention to the interval $[0, 1]$.} which is also publicly known to the agent, and can choose a signaling mechanism 
$\{\messageprob_i(\cdot \condition x), \messagespace_i\}$,
where $\messagespace_i$ is a signal space and 
$\messageprob_i(\signal \condition x) \in [0, 1]$
specifies the conditional distribution of signal $\signal \in \messagespace_i$ when the prize value $x \sim \priorCDF$ is realized.
The senders' signaling mechanisms $\{\messageprob_i(\cdot \condition x), \messagespace_i\}_{i\in[n]}$ are known to the agent in advance.

Given the boxes' signaling mechanisms, 
the agent can learn about the boxes' prize values by inspecting the boxes with paying the inspection cost $c_i > 0$ and observing their signal realizations in sequence. 
When the agent inspects box $i$, she only observes a realized signal $q$ drawn according to the conditional distribution $\messageprob_i$. 
After observing the signal realization of box $i$,
the agent updates her prior to a posterior distribution about the underlying prize value $x_i$ of box $i$.
Importantly, the agent can observe the realized prize value $x_i$ only if she stops the search and chooses to take the box $i$ among which have already been inspected.
%
%
The agent's goal is to determine a strategy $\pi$ to 
inspect boxes to maximize her expected payoff in \eqref{agent_exp_payoff}. 

In our setting, each box $i$ (a.k.a., sender $i$) 
is competing with each other for the final selection from the agent. Specifically, the payoff of each sender $i$ can be expressed as $\indicator{\selectIdx_i = 1}$.
Namely, a sender obtains payoff $1$ if he is selected and payoff $0$ if he is not selected.\footnote{ Our results generalize immediately to settings where each sender $i$ prefers being chosen over not being chosen since any such case leads to the same ultimate  objective of maximizing $\Pr( \indicator{\selectIdx_i = 1})$.}
\wtrTE{\xhdr{Solution concept} 
The timing of our competitive information design 
game can be detailed as follows:
First, each sender commits to an information strategy (a.k.a., a signaling mechanism).
Second, the agent observes all boxes' \infoStrategies, 
and uses an inspection strategy $\pi$ to determine 
how to inspect and when to stop the inspection.
Finally, the agent observes signal realizations among all inspected boxes, and decides which box to take the prize value. 
When the agent is indifferent 
between multiple boxes, she chooses 
one of them uniformly at random.

The main focus of this paper is how the boxes design the signal mechanisms and how these mechanisms subsequently shape the agent's search and stopping behavior, ultimately affecting the box's own payoff.
To this end, we focus on subgame-perfect Nash equilibrium solution concept. 
When it comes to analyze the Nash equilibrium among boxes' game, we assume all boxes are {\em ex ante} symmetric in the sense that they have the identical prior prize distribution $\priorCDF \equiv \priorCDF_i$ and they have the same inspection cost $c \equiv c_i$.
Given such {\em ex ante} symmetry,
we thus follow the earlier works~\citep{GK-16,GK-17,AW-23} and focus on the solution concept of pure-strategy equilibria among boxes' game.}\footnote{
We will use equilibrium synonymously with 
pure symmetric equilibrium.} 


\section{Characterizing Agent's Optimal Strategy}
\label{sec-opt-searching}

When the senders are not strategically revealing the prize value information, the agent's optimal inspection and the stopping rule can be characterized by an elegant threshold-based strategy proposed by \cite{wei-79}.
We below describe this threshold-based strategy which will be useful for our subsequent analysis.
In particular, \cite{wei-79} defines a notion of \emph{reservation value} for the corresponding prize value distribution:
\begin{definition}[Reservation Value]
\label{sigma_defn}
For a box with prize distribution $H \in \Delta([0, 1])$ and a cost $c > 0$ for inspecting this box, the value $\sigma_H$ satisfying $\sigma_H = \sup\{\sigma: \expect[x\sim H]{\max\{x - \sigma,0\}} = c\}$ is referred to as the \emph{reservation value}.  
\end{definition}
With the notion of reservation value, 
the agent's optimal strategy can be characterized by the simple procedure below. 
\begin{theorem}[\citealp{wei-79}]
\label{thm_optimal_searching}
Let $\sigma_{\priorCDF_i}$ be the reservation value of box $i$'s prize distribution $\priorCDF_i$ with inspection cost $c_i$. Then the following  strategy is optimal for the agent: the agent  (i) inspects each box in order of decreasing  $\{\sigma_{\priorCDF_i}\}_{i\in[n]}$; (ii) stops when the largest observed prize value exceeds the next uninspected reservation value and selects box that has the largest observed prize value.
\end{theorem} 
When each sender that is associated with a box strategically reveals information about the prize value, the agent's optimal search and stopping strategy becomes less obvious.
Perhaps surprisingly, we show that the agent's optimal strategy still closely resembles the structure outlined in \Cref{thm_optimal_searching}.
The main differences here are: (i) instead of computing the reservation values based on the prior prize distributions $\{\priorCDF_i\}_{i\in[n]}$, the agent now computes these values based on the posterior mean distributions induced from the boxes' signaling mechanisms; (ii) the agent determines whether to stop the search based on the mean of the posterior belief induced from the observed signal and the subsequent reservation value. 
\begin{theorem}
\label{thm:agent opt}
Given boxes' signaling mechanisms $\{\messageprob_i(\cdot \condition x), \messagespace_i\}_{i\in[n]}$, let $\{G_i\}_{i\in[n]}$ where each $G_i\in\Delta([0, 1])$ is the corresponding {\em distribution of posterior means} induced from the mechanism $\{\messageprob_i(\cdot \condition x), \messagespace_i\}$.
Then under the mechanisms $\{\messageprob_i(\cdot \condition x), \messagespace_i\}_{i\in[n]}$, the agent's optimal strategy is:
(i) computing the reservation value $\sigma_{G_i}$ for each box $i$ based on $G_i$, and inspecting boxes in decreasing order of $\{\sigma_{G_i}\}_{i\in[n]}$; 
(ii) when inspecting box $i$, she observes a signal realization, computes prize posterior mean and stops if the largest posterior mean exceeds the next uninspected reservation value. If she stops, she selects box that has the largest posterior mean to take the prize value.
\end{theorem}
To understand the intuition behind \Cref{thm:agent opt}, notice that whenever the agent decides to stop searching, she always chooses the box with the highest expected prize value based on her posterior beliefs. 
This follows from the fact that the agent is risk-neutral and has a unit demand.
Additionally, when deciding which box to inspect next, the agent bases her decision on the posterior beliefs formed from previously inspected boxes, along with the prior prize distributions and signaling mechanisms of the remaining boxes.
Since the agent is risk-neutral, only the posterior means of the previously inspected boxes influence her subsequent search and stopping decisions. These insights are rigorously formalized in the proof of \Cref{thm:agent opt}.

As a consequence of \Cref{thm:agent opt}, the agent's decisions in her optimal search and stopping strategy depend on the boxes' signaling mechanism (and the realized signals) through the induced distributions of posterior means (and the realized posterior means).
In other words, the box $i$'s expected payoff remains unchanged if we consider a new game in which we replace box $i$'s signaling mechanism with its corresponding posterior mean distribution.
While this is a direct consequence of \Cref{thm:agent opt}, we summarize this equivalence of strategies as follows.
\begin{corollary}
\label{cor:posterior mean dist equivalence}
Fix all other boxes' signaling mechanisms and under the agent's optimal strategy defined in \Cref{thm:agent opt}, the box $i$'s expected payoff would remain the same if we consider a new game in which we replace box $i$'s signaling mechanism with its induced distribution of posterior means.
\end{corollary}
We would like to highlight that, for tractability, reformulating the sender's signaling mechanism as a distribution of posterior means is a common approach in previous information design literature. 
This reformulation is indeed without of loss of generality when the sender's payoff depends only on the expected state. This occurs when the receiver's optimal action depends only on the expectation of the posterior belief and sender's preferences over receiver's actions do not depend on the realized state.
However, this reformulation is not generally without of loss of generality without \Cref{thm:agent opt}. 
For example, the agent could adopt a strategy based on additional distributional information, such as the variance of the posterior belief. 
In such case, the boxes'  payoff would depend on more than just the posterior mean.
Nevertheless, \Cref{thm:agent opt} allows us to establish this strategy equivalence by leveraging the structure of the agent's optimal strategy.
Reformulating the boxes' signaling mechanisms as the distributions of posterior means significantly simplify our subsequent analysis.
\cite{AW-23} have studied a similar search setting to ours while they focus on binary prize value.
It is worthy of note that, unlike our setting, when the prize takes bianry value, this reformulation is straightforward and valid regardless of the agent's strategy.

With \Cref{cor:posterior mean dist equivalence}, 
a natural next question is which distributions over posterior means can be implemented by some signaling mechanisms given the prior prize distribution.
This question can be answered using the notion of mean-preserving spread (MPS), which characterizes the \emph{feasible}  distributions that can represent the sender's information strategies.
\begin{definition}[Mean-preserving Spread]
\label{MPS_defn}
A distribution $H\in\Delta([0, 1])$ 
is a Mean-preserving Spread (MPS) of a distribution $G \in\Delta([0, 1])$,
represented as $H \blackwellOrder G$,
if and only if for all $\sigma\in[0, 1],\int_0^\sigma H(x)dx 
\ge \int_0^\sigma G(x)dx$
where the inequality holds as equality for $\sigma = 0$.
\end{definition}
It turns out that a distribution $G$ over posterior means can be induced by some signaling mechanism from prior prize distribution $\priorCDF$ if and only if $\priorCDF$ is an MPS of $G$. 

\begin{lemma}
[\citealp{Blackwell-79}]
\label{lem_MPS_singaling}
There exists a signaling mechanism that induces the distribution $G$ over posterior means \hfr{from prior prize distribution $\priorCDF$} if and only if $\priorCDF \blackwellOrder G$.
\end{lemma}
With \Cref{lem_MPS_singaling} and \Cref{thm:agent opt},  
we can without loss of generality assume that each box $i$'s strategy\footnote{
We will use sender's strategy synonymously 
with sender's information strategy.} 
is to directly choose a distribution $G_i \in \Delta([0, 1])$
that satisfies $G_i: \priorCDF_i \blackwellOrder G_i$, 
without the need of concerning the design of the underlying signaling mechanism $\{\messageprob_i(\cdot \condition x), \messagespace_i\}$.
In the following discussion, we directly refer to $G_i$ as sender $i$'s \infoStrategy.
Moreover, following Blackwell's ordering of informativeness~\citep{Blackwell-53},
we say a \infoStrategy\ $G'$ is \emph{more informative} 
than $G$ if $G'$ is an MPS of $G$, i.e., $G' \blackwellOrder G$.


\section{Equilibrium Analysis}
\label{sec-continuous}

\newcommand{\cH}{\mathcal{H}}

In this section, we characterize the equilibrium for the senders' game of competitive information design in a symmetric environment, namely,
$H \equiv H_i, c \equiv c_i, \forall i\in[n]$.
In particular, we give sufficient and necessary
conditions of the existence of pure symmetric equilibrium.
We also characterize the unique equilibrium strategy
if the pure symmetric equilibrium exists.

\begin{figure}[h]
\centering
\subfloat[
\label{Fig: ex_no_Nash_a}
]{
\definecolor{battleshipgrey}{rgb}{0.52, 0.52, 0.51}

\begin{tikzpicture}
\begin{axis}[
axis line style=gray,
axis lines=middle,
xtick style={draw=none},
ytick style={draw=none},
xticklabels=\empty,
yticklabels=\empty,
xmin=-0.05,xmax=1,ymin=-0.15,ymax=1.1,
width=0.45\textwidth,
height=0.45\textwidth,
samples=200]

\addplot[domain=0:1, gray!40!white, line width=2mm] (x, {x^(0.3)});

\addplot[domain=0:0.1122,      gray!95!white,line width=1mm] (x, {0.2431^(0.3)/0.1122 * x});
\addplot[domain=0.1122:0.5613, gray!95!white,line width=1mm] (x, {0.2431^(0.3)});
\addplot[domain=0.5613:1,      gray!95!white,line width=1mm] (x, {1});

\addplot [black, line width = 0.3mm, densely dashed] coordinates {(0,0) (0.2308, 0)};
\addplot [black, line width = 0.3mm, densely dashed] coordinates {(0.2308, 1) (1, 1)};

\addplot[gray, dotted] coordinates {(0.1122,0.654238) (0.1122,0)};
\addplot[gray, thick] coordinates {(0.1122,0.01)
(0.1122,-0.01)};
\addplot[] coordinates {(0.1122,0)} node[below, pos=1, yshift = -0.14cm]{$\bar{x}_G$};

\addplot[gray, dotted] coordinates {(0.2431,0.654238) (0.2431,0)};
\addplot[gray, thick] coordinates {(0.2431,0.01)
(0.2431,-0.01)};
\addplot[] coordinates {(0.2431,0.)} node[below, pos=1, yshift = -0.15cm]{$\sigma_H$};

\end{axis}

\end{tikzpicture}
}\qquad
\subfloat[
\label{Fig: ex_no_Nash_b}
]{
\definecolor{battleshipgrey}{rgb}{0.52, 0.52, 0.51}

\begin{tikzpicture}
\begin{axis}[
axis line style=gray,
axis lines=middle,
xtick style={draw=none},
ytick style={draw=none},
xticklabels=\empty,
yticklabels=\empty,
xmin=-0.05,xmax=1,ymin=-0.15,ymax=1.1,
width=0.45\textwidth,
height=0.45\textwidth,
samples=200]

\addplot[domain=0.001:0.9999, gray!40!white, line width=2mm] (x, {1/(1+(x/(1-x))^(-3))});

\addplot[domain=0:0.4463,      gray!95!white,line width=1mm] (x, {1/(1+(x/(1-x))^(-3))});
\addplot[domain=0.4463:0.6570631198369846, gray!95!white,line width=1mm] 
(x, {2.73839136526 * x-0.8784544738193616});
\addplot[domain=0.6570631198369846:0.756923932887,      gray!95!white,line width=1mm] (x, {0.9208414999730338});
\addplot[domain=0.756923932887:1,      gray!95!white,line width=1mm] (x, {1});

\addplot[domain=0:0.596067878321, black, densely dashed, 
line width=0.3mm] (x, {1/(1+(x/(1-x))^(-3))});
\addplot[domain=0.596067878321:0.678124753117, black, densely dashed, 
line width=0.3mm] (x, {0.7626608138636253});
\addplot[domain=0.678124753117:1, black, densely dashed, 
line width=0.3mm] (x, {1});

\addplot[gray, dotted] coordinates {(0.596067878321,0.7626608138636253) (0.596067878321,0)};
\addplot[gray, thick] coordinates {(0.596067878321,0.01)
(0.596067878321,-0.01)};
\addplot[] coordinates {(0.596067878321,0.)} node[below, pos=1, xshift = -0.2cm]{$x^\dagger$};

\addplot[gray, dotted] coordinates {(0.6570631198416026,0.9208414999730338) (0.6570631198416026,0)};
\addplot[gray, thick] coordinates {(0.6570631198416026,0.01)
(0.6570631198416026,-0.01)};
\addplot[] coordinates {(0.6570631198416026,0.)} node[below, xshift = 0.02cm, yshift=-0.1 cm]{$\bar{x}_G$};

\addplot[gray, dotted] coordinates {(0.6938,0.9208414999730338) (0.6938,0)};
\addplot[gray, thick] coordinates {(0.6938,0.01)
(0.6938,-0.01)};
\addplot[] coordinates {(0.6938,0.)} node[below, xshift = 0.4cm, yshift=-0.15cm]{$\sigma_H$};

\end{axis}

\end{tikzpicture}
}
\caption{In both figures, the prior $H$ is the gray solid line, the distribution $G$
that satisfies conditions $(i)$--$(ii)$
in \Cref{thm_nash_iff} is the deep gray solid line. 
The profitable deviation $F$ is then the 
black dashed line. 
See the detailed descriptions in \Cref{ex_no_Nash_a}
and \Cref{ex_no_Nash_b}.
(a): 
Equilibrium does not exist as it violates the 
the case $(a)$ in condition $(iii)$.
(b): 
Equilibrium does not exist as it violates 
the case $(b)$ in condition $(iii)$.
\label{ex_no_nash}
}
\end{figure}
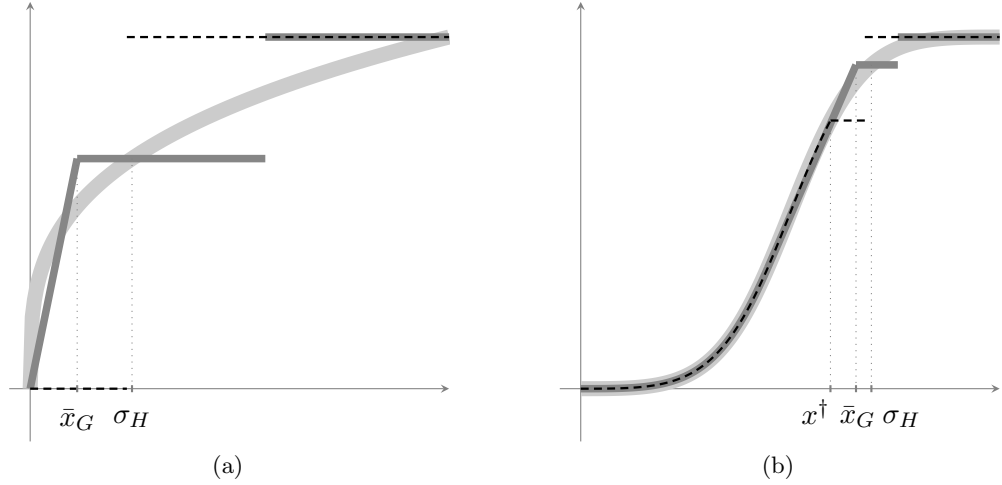

Before stating our main results, we first define a special structure of senders' \infoStrategies\ that will be useful to 
help describe the structure of the equilibrium strategy. 
\begin{definition}
[Alternating $(n-1)$-linear MPS -- \citet{HKB-19}]
\label{defn_linear_MPS}
Given a prior $H$, 
$G$ exhibits alternating $(n-1)$-linear MPS behavior
in the interval $[a, b]$
if whenever $G$ is not 
fully revealing information
in a subinterval $[x_1, x_2] \subseteq [a, b]$, 
$G^{n-1}$ 
is linear over  $[x_1, 
\min\{x_{2}, \max_{x\in[a, b]}\{x: x\in\supp{G}\}\}]$
and $H \blackwellOrder_{[x_1, x_{2}]} G$.
\end{definition}
With the above structure, our main result in this section
can be stated as follows.
\begin{theorem}
\label{thm_nash_iff}
For any prior $H$ and any cost $c \ge 0$,
given a strategy $G$ and its
$\bar{x}_G := \max\{x\in[0, \sigma_H]: x\in\supp{G}\}$,
$(G, \ldots, G)$ is an equilibrium
if and only if
\begin{enumerate}
    \item[(i)] 
    $\sigma_G = \sigma_H$;
    \item[(ii)] 
    $G^{n-1}$ is convex over $[0, \bar{x}_G]$ and $G$ exhibits alternating $(n-1)$-linear MPS behavior over $[0, \sigma_H]$;
    \item[(iii)] 
    deviating
    to a strategy $F$ where $\sigma_F = \max\{\bar{x}_G, \lambda - c\}$ is not profitable.
    More concretely, 
    \begin{enumerate}
        \item[(a)] if $\lambda - c \ge \bar{x}_G$, the optimal  value from deviation
        $G(\lambda - c)^{n-1} $ is at most $ \sfrac{1}{n}$;
        \item[(b)] if $\lambda - c < \bar{x}_G$, the optimal deviation value
        $\int_0^{x^\dagger} G(x)^{n-1}dH(x) 
        + H(\sigma_H)^{n-1}(1 - H(x^\dagger)) \le \sfrac{1}{n}$
        where $x^\dagger$ uniquely satisfies 
        $\int_{x^\dagger}^1 (x - \bar{x}_G)dH(x)=c$.
    \end{enumerate}
\end{enumerate}
\end{theorem}
We interpret and examine each condition in the theorem below. 
Condition $(i)$ indicates that the reservation
value of the equilibrium strategy $G$ must achieve
its maximum, i.e., $\sigma_G = \sigma_H$ 
(recall that from \Cref{interval_MPS}, 
we know $\sigma_H$ is the maximum reservation value 
that is attainable for any 
feasible information strategy of prior $H$).
This aligns with the intuition that 
each sender prefers to be inspected earlier rather than later. 
Condition $(ii)$ characterizes the structure of feasible equilibrium strategy under the reservation value $\sigma_H$.
As we elaborate shortly, the first two 
conditions can uniquely\footnote{
The uniqueness here means the behavior of $G$ 
over $[0, \sigma_H]$ is unique. 
Note that \Cref{thm_nash_iff}
only states the conditions for the support of $G$ 
that is in $[0, \sigma_H]$.
Indeed, one can show
that if $(G, \ldots, G)$ is an equilibrium, 
then $(G, \ldots, G', \ldots, G)$ is also an 
equilibrium as long as $G'(x) = G(x), 
\forall x\in[0, \sigma_H]$. The reason is that once
we pin down the reservation value of all senders' strategies
to be $\sigma_H$, each sender's expected payoff only
depends on the behavior of his strategy in $[0, \sigma_H]$
(see \Cref{equilibrium_simplication} 
for detailed discussions). }
pin down a distribution $G$. 
Lastly, condition $(iii)$ 
verifies whether $G$ that satisfies the first two conditions is indeed
an equilibrium strategy. 
Essentially, 
there are only two scenarios:
(a) If $\lambda - c \ge \bar{x}_G$,  
deviating to no information strategy for a sender
is the most profitable.
(b) If $\lambda - c < \bar{x}_G$, 
deviating to a 
strategy $F$ that has the reservation value $\bar{x}_G$ and 
satisfies $F(x) = H(x), \forall x\le x^\dagger$
and has no support between $x^\dagger$ and $\bar{x}_G$
is the most profitable
(we give an example on how to compute such deviation $F$,
see the blue dotted line in \Cref{Fig: ex_no_Nash_b}). 
\footnote{This specific structure of $F$
is largely due to the convexity of $G^{n-1}$
over $[0, \bar{x}_G]$, it will be proved in \Cref{opt_deviation_structure}.}
In either case, the optimal deviation value 
can be computed in a closed form, 
so we can verify whether $G$ is indeed an equilibrium strategy.

Note that in the special case where the inspection cost $c=0$,
our problem reduces to a simpler setting, in which 
the agent does not need to choose which senders to inspect and in what order 
as he can inspect all senders for free. 
In this setting, 
\citet{HKB-19}\footnote{
In their model, the agent firstly observes
all realized prize values $\{x_i\}_{i\in[n]}$, and then
selects the sender that has the maximum value.
This is equivalent to our setting 
with $c = 0$. To see this, note 
when $c = 0$, the reservation value of 
any \infoStrategy\ goes to infinity. 
Thus, though the agent sequentially inspects senders, 
he would inspect all senders and select the best one.\label{special-case}
} 
show that there always
exists a unique \hfr{symmetric equilibrium strategy $(G, \cdots, G)$  satifying that} $G^{n-1}$ is convex over the support of $G$, and $G$ exhibits the above alternating behavior over $[0, 1]$ as defined in \Cref{defn_linear_MPS}. 
Our result strictly generalizes their result.
First, we can see that our conditions $(i)$--$(iii)$ 
are always satisfied when $c = 0$:
When there is no inspection cost,
both $\sigma_G$ and $\sigma_H$ approach $+\infty$.
For our condition $(ii)$, $G$ exhibiting
alternating behavior over $[0, \sigma_H]$
is equivalent to exhibiting
alternating behavior over $[0, 1]$.
For condition $(iii)$, given a distribution $G$
satisfying condition $(ii)$ over $[0, 1]$, 
we always have $\lambda - c = \lambda < \bar{x}_G$
as $G$ has no support over $[\bar{x}_G, 1]$.
When $c = 0$, we have $x^\dagger = 1$, 
and
$\int_0^{x^\dagger} G(x)^{n-1}dH(x) + 
H(\sigma_H)^{n-1}(1 - H(x^\dagger))
= \int_0^{1} G(x)^{n-1}dH(x) 
\le \sfrac{1}{n}$ holds for sure. 
To see this, 
note that \citet{HKB-19} have showed that such
$G$ is the equilibrium strategy when $c = 0$.
Thus, by definition, we have 
$\int_0^1 G(x)^{n-1} dH(x)\le \int_0^1 G(x)^{n-1} dG(x)
= \sfrac{1}{n}$.

When inspection cost $c>0$, 
a pure symmetric equilibrium 
might not exist.
We present two examples (see \Cref{ex_no_nash}) 
where the 
pure symmetric 
equilibrium does not exist.
Each of the examples violates one of the cases 
in condition $(iii)$.

\begin{example}[\hfr{ Equilibrium Non-Existence: Violation of case $(a)$ in condition $(iii)$}]
\label{ex_no_Nash_a}
Consider prior $H(x) = x^{0.3}$ 
(the gray solid line in \Cref{Fig: ex_no_Nash_a}), 
$n=2$, and a cost $c = 0.11$. 
With this prior, one can compute
$\lambda = 0.2308$, $\sigma_H = 0.2431$, 
and $\sigma_{\NI} = \lambda - c = 0.1208$. 
Using the conditions $(i)$--$(ii)$ in 
\Cref{thm_nash_iff}, one can 
compute a unique $G$ with $\bar{x}_{G} = 0.1122$
(notice that the CDF $G$ is linear over $[0, \bar{x}_{G}]$ 
and $\bar{x}_{G}$ is the unique solution satisfying: 
$\frac{1}{2} \bar{x}_G H(\sigma_H) + (\sigma_H - \bar{x}_G) H(\sigma_H) = \int_0^{\sigma_H} H(x)dx$). 
However, such $G$ is not an equilibrium strategy 
as one can deviate to a 
No information disclosure strategy $G_{\NI}$ 
to achieve a higher payoff 
$G(\sigma_\NI) = 0.6542 > 0.5$.
\end{example}

\begin{example}[\hfr{ Equilibrium Non-Existence: Violation of case $(b)$ in condition $(iii)$}]
\label{ex_no_Nash_b}
Consider prior
$H(x) = \frac{1}{1+\left(\frac{x}{1-x}\right)^{-3}}$
(the gray solid line in \Cref{Fig: ex_no_Nash_b}),
$n = 2$, and a cost $c = 0.005$.
With this prior, one can compute
$\lambda = 0.5$, and $\sigma_H = 0.6938$.
Using the conditions $(i)$--$(ii)$ in 
\Cref{thm_nash_iff}, one can 
compute a unique $G$ where $\bar{x}_G = 0.6571$
(notice that from the conditions $(i)$--$(ii)$, 
the distribution $G$ first equals to the prior $H$
over $[0, x']$ for some point $x'$, then is linear over $[x', \bar{x}_G]$, and furthermore, the slope of the linear part exactly equals to $H'(x')$. Thus, $\bar{x}_G$ and $x'$ is the unique solution 
satisfying $\bar{x}_G = x' + \frac{H(\sigma_H) - H(x')}{H'(x')}$; $\int_0^{x'} H(x)dx + \frac{H(x') + H(\bar{x}_G)}{2} (\bar{x}_G - x') + (\sigma_H - \bar{x}_G) H(\sigma_H) = \int_0^{\sigma_H}H(x)dx$). 
However, such $G$ is not an equilibrium as one can 
deviate to a strategy 
$F$ with $x^\dagger = 0.5961$ to a higher payoff $0.5048 > 0.5$. 
$F$ has reservation value $\sigma_F = \bar{x}_G$,
and $F(x) = H(x), \forall x\in[0, x^\dagger]$, 
and $F$ has no support over $[x^
\dagger, \bar{x}_G]$.
\end{example}

\subsection{Proof of \texorpdfstring{\Cref{thm_nash_iff}}{}}
\label{subsec-proof-main}
In this section, we present our proof for \Cref{thm_nash_iff}.

\xhdr{Technical challenges and proof overview} 
Determining whether a particular
strategy profile $(G, \ldots, G)$ is 
a symmetric equilibrium can be challenging, 
as it depends on the full set $\cH$ of feasible \infoStrategies,
i.e., $\cH := \{F: H\blackwellOrder F\}$, 
that each sender can deviate to.
When the agent uses the optimal inspection 
strategy, 
however, using the observation we obtain in
\Cref{prop_informativness_sigma}, 
one can first show that a \infoStrategy\ $G$ 
can be an equilibrium \infoStrategy\ only if 
it satisfies $\sigma_G = \sigma_H$.
This observation shrinks the set that contains 
any possible equilibrium strategy to the set
$\cH(\sigma_H) 
:= \{F: H\blackwellOrder F \wedge \sigma_F = \sigma_H\}$.
Next, using the conditions provided in 
\Cref{interval_MPS}, and examining the fixed point 
problem over the set 
$\cH(\sigma_H)$,
we can uniquely pin down the behavior of $G$
over the interval $[0, \sigma_H]$
if $G$ is the equilibrium strategy.

The above procedure helps us pin down the necessary conditions for $G$ to be the equilibrium strategy.
To verify whether the identified $G$ is indeed
the equilibrium strategy, we need to show that
no sender has profitable deviation under the 
strategy profile $(G, \ldots, G)$. 
This step is challenging since we again need to 
examine all possible deviations that one sender 
can deviate to when all other senders use strategy $G$. 
Different deviation \infoStrategies\ 
have different reservation values, which impact
the order that the agent inspects the boxes, 
and subsequently change the deviation payoff.
In a more detail, when a deviation \infoStrategy\ $F$
has reservation value 
$\sigma_F = \sigma <\sigma_H$,
let  $\senderURealize(x)$  be
the sender's deviation payoff
as a function of the realized value $x \sim F$, 
it can be shown that
$\senderURealize(x) = 
\min\{G(x)^{n-1}, G(\sigma)^{n-1}\}$, in which 
the shape of $\senderURealize(\cdot)$ 
depends on the choice of $\sigma$. 
Thus, there is no single program that 
can encode sender's deviation problem.
Instead, 
our solution is that, for every possible 
reservation value $\sigma$, we consider the corresponding
linear program (note that the constraint $\sigma_F = \sigma$
can be formulated as a linear constraint), 
and then characterize its optimal 
deviation \infoStrategy. 
We then show that the optimal deviation value 
is single-peaked (with 
the peak at $\sigma^* := \max\{\sigma_{\NI}, \bar{x}_G\}$)
w.r.t. $\sigma\in[\sigma_\NI, \sigma_H)$.
To this end, to account for the additional constraint 
$\sigma_F = \sigma$, we extend the 
verification tool provided in \citet{DM-19}
to show what the optimal dual solution
must look like, and then show there exists an 
optimal primal solution that satisfies complementary slackness.

To summarize, the analysis mainly 
consists of following steps:
\begin{itemize}
    \item \textbf{Step 1}. 
    In this step, we prove the 
    condition $(i)$ in \Cref{thm_nash_iff},
    namely, for any prior $H$, 
    if there exists a symmetric equilibrium $(G, \ldots, G)$, 
    it must be that $\sigma_G = \sigma_H$ 
    (see \Cref{max_sigma_nash}).
    
    \item \textbf{Step 2}. 
    In this step, we show that 
    no sender has profitable deviation 
    to a strategy $F\in\cH(\sigma_H)$
    if all other senders 
    use strategy satisfying
    conditions $(i)$--$(ii)$ in \Cref{thm_nash_iff}
    (see \Cref{symme_deviation}).
    
    \item \textbf{Step 3}.  
    In this step, we show that 
    when all other senders 
    use strategy $G$ satisfying
    conditions $(i)$--$(ii)$ in \Cref{thm_nash_iff},
    then no sender has profitable deviation
    if and only if 
    condition $(iii)$ holds
    (see \Cref{lem_most_profitable_devia}).
\end{itemize}
Below, we first provide detailed analysis of the above steps.
The proof of the main result \Cref{thm_nash_iff}
follows from combining
the results of these steps.

\xhdr{Step 1 -- Characterizing the reservation value of equilibrium strategy}
\begin{restatable}{lemma}{maxsigmanash}
\label{max_sigma_nash}
For any $H$, if there exists a symmetric equilibrium 
$(G, \ldots, G)$, 
it must be that $\sigma_G = \sigma_H$.
Each sender's expected payoff is $\sfrac{1}{n}$ 
at any \hfr{symmetric} equilibrium.
\end{restatable}
\ifTE
\else
\begin{proof}[Proof Sketch of \Cref{max_sigma_nash}]
Given any symmetric strategy $(G, \ldots, G)$ 
where $\sigma_G < \sigma_H$, 
each sender $i$'s expected payoff $\senderU_i(G, \ldots, G)$
can be expressed as 
\begin{align*}
    \senderU_i(G, \ldots, G) 
    := \prob{\selectIdx_i = 1|\openIdx_i = 1} 
    \cdot \prob{\openIdx_i = 1}~,
\end{align*}
where $\prob{\openIdx_i = 1}$ is the probability of 
sender $i$ being inspected by the agent and 
$\prob{\selectIdx_i = 1|\openIdx_i = 1}$ is 
the expected payoff conditional on being inspected.
As there always exists probability such that sender $i$ 
is never inspected by the agent, we have 
\begin{align*}
    \prob{\openIdx_i = 1} \equiv 1 - \delta < 1~.
\end{align*}
Now let $U_i^S(x)$ denote the sender $i$'s expected payoff 
conditional on being inspected 
and the value $x$ realizing. 
Then we have
\begin{align*}
    \prob{\selectIdx_i = 1|\openIdx_i = 1} = \int_0^1 U_i^S(x)dG(x)~.
\end{align*}
Now let $F: H\blackwellOrder F$ be a strategy 
satisfying $\sigma_F > \sigma_G$ and also
\begin{align*}
    \int_0^1 U_i^S(x)dF(x) > \int_0^1 U_i^S(x)dG(x) - \varepsilon~,
\end{align*}
for a small $\varepsilon > 0$.
Note as $\sigma_G < \sigma_H$, such $F$ must exist
(we defer the detailed construction
of such $F$ to the \Cref{proof-sec-continuous}).
Then by deviating to strategy $F$, from \Cref{prop_informativness_sigma},
we know sender $i$'s probability of 
being inspected is increased to $1$. 
Thus, 
\begin{align*}
    \senderU_i(G, \ldots, F, \ldots, G) - \senderU_i(G, \ldots, G) 
    & > \int_0^1 U_i^S(x)dF(x) -  \int_0^1 U_i^S(x)dG(x) \cdot (1 - \delta)\\
    & = \delta \cdot \int_0^1 U_i^S(x)dG(x) - \varepsilon > 0~,
\end{align*}
where the last inequality is by 
choosing a sufficiently small $\varepsilon$.
As a result, such deviation is profitable. 

Clearly, each sender's expected payoff is $\sfrac{1}{n}$ 
at any equilibrium. 
Suppose not, then the sender who has expected payoff 
smaller than $\sfrac{1}{n}$ 
can improve his expected payoff by simply mimicking
another sender's strategy who has higher payoff
than
$\sfrac{1}{n}$ . 
\end{proof}
\fi

\xhdr{Step 2 -- Characterizing the behavior
of $G$ over the interval $[0, \sigma_H]$}
Now we use the result in \Cref{lem_same_density}
and the characterization in \Cref{interval_MPS}
to prove the condition $(ii)$.
\begin{lemma}
\label{symme_deviation}
Given prior $H$,
under the strategy profile $(G, \ldots, G)$
where $G$ satisfies 
the conditions $(i)$--$(ii)$ in \Cref{thm_nash_iff},
then no sender has a profitable deviation 
to a strategy $F$ where $\sigma_F = \sigma_H$.
Meanwhile, if $(G, \ldots, G)$ is an 
equilibrium, then the behavior of $G$ over
the interval $[0, \sigma_H]$
must satisfy the condition $(ii)$ in \Cref{thm_nash_iff}. 
\end{lemma}
The intuition behind the proof for the above result 
is as follows. 
Given all other senders using strategy $G$ and sender $i$
using strategy $F$ where $\sigma_F = \sigma_H$, 
with the result in \Cref{lem_same_density}, it can be 
shown that sender $i$'s expected payoff only depends on 
the behavior of $F$ over the interval $[0, \sigma_H]$.
Then using the characterization in \Cref{interval_MPS}, 
and the earlier results in \citet{HKB-19}, 
we show sender $i$'s best deviation in the set $\cH(\sigma_H)$
is indeed $G$ itself.

\xhdr{Step 3 -- Verifying whether $G$ is indeed an 
equilibrium strategy}
Now to argue whether $G$, which satisfies
the conditions $(i)$--$(ii)$
in \Cref{thm_nash_iff},
is an equilibrium strategy,
it remains to show that 
no sender has a profitable deviation 
to a strategy $F$ 
that has $\sigma_F < \sigma_H$ 
if all other senders 
use the strategy $G$.
In other words, 
we need to show that whenever 
we fix a $\sigma \in [\sigma_{\NI}, \sigma_H)$, 
the best payoff for a
sender $i$ to deviate to a strategy 
$F\in\cH(\sigma) 
:= \{F: H\blackwellOrder F \wedge \sigma_F = \sigma\}$ 
is no larger than $\sfrac{1}{n}$. 
Given sender $i$ using $F$ where $\sigma_F = \sigma < \sigma_H$, 
and other senders using $G$, we have
\ifTE
\else
\begin{align*}
\senderU_i(G, \ldots, F, \ldots, G) 
& = G(\sigma)^{n-1} \cdot \int_{\sigma}^1 dF(x) 
+ \int_0^\sigma G(x)^{n-1}dF(x)~.
\end{align*}
Using integral by parts and rearranging the terms, 
we can get
\fi
\begin{align}
    \label{deviation_payoff_smaller_RV}
    \senderU_i(G, \ldots, F, \ldots, G)  
    = \int_0^1 \min\left\{G(x)^{n-1}, G(\sigma)^{n-1}\right\} dF(x)~.
\end{align}
The proof of \Cref{symme_deviation}
and the above payoff deviation  
have following implication that only the 
behavior over the interval $[0, \sigma_H]$ of the strategy $G$ matters for the equilibrium.

\begin{restatable}{corollary}{equilibriumsimplication}
\label{equilibrium_simplication}
Given a prior $H$, if $(G, \ldots, G)$
is an equilibrium, 
then the strategy profile $(G_1, \ldots, G_n)$
where $ \forall i, G_i(x) = G(x), \forall x\in[0, \sigma_H]$
is also the equilibrium.
\end{restatable}
Fix a $\sigma \in [\sigma_{\NI}, \sigma_H)$, we now consider
following sender $i$'s best response strategy that is 
subject to the constraint $\sigma_F = \sigma$ 
\begin{align}
    \label{new_program_general}
    \max_{F\in\cH(\sigma)}   
    \int_0^1 \min\left\{G(x)^{n-1}, G(\sigma)^{n-1}\right\} dF(x)~.
\end{align}
Given $\sigma$, 
let $\OPT_\sigma$ denote the optimal value
of the above program.
Essentially, $G$ is equilibrium strategy
must satisfy that
\begin{align}
    \max_{\sigma: \sigma\in[\sigma_\NI, \sigma_H)}
    \OPT_\sigma \le \frac{1}{n}~.
    \label{deviation_goal}
\end{align}
In below analysis, we characterize
the most profitable deviation given 
all other senders using strategy $G$. 
In particular, to guarantee~\eqref{deviation_goal},
we show that, depending
on the relative 
value $\sigma_\NI$ and $\bar{x}_G$,
it suffices
to only consider one deviation:
either deviating to 
no information disclosure strategy 
(if $\sigma_\NI > \bar{x}_G$)
or deviating to a strategy whose reservation 
value equals to $\bar{x}_G$
(if $\sigma_\NI \le \bar{x}_G$).
\begin{lemma}
\label{lem_most_profitable_devia}
Fix a prior $H$ and the cost $c > 0$, 
given all other senders using $G$ 
that meets the 
conditions $(i)$--$(ii)$ 
in \Cref{thm_nash_iff}, then 
\begin{enumerate}
    \item[(a)]
    if $\sigma_{\NI} = \lambda - c > \bar{x}_G$, the most profitable 
    deviation is \noInforStrategy;
    
    \item[(b)]
    if $\sigma_{\NI} = \lambda - c \le \bar{x}_G$,
    the most profitable 
    deviation is a 
    strategy $F$ where $\sigma_F = \bar{x}_G$.
\end{enumerate}
\end{lemma}
The condition $(iii)$ in \Cref{thm_nash_iff} simply follows 
by ensuring that the value of most profitable 
deviation is no larger than $\sfrac{1}{n}$.
To prove \Cref{lem_most_profitable_devia}, 
for the case $\sigma_{\NI} \le \bar{x}_G$,
we separate our discussions in two regimes:
for $\sigma\in[\sigma_{\NI}, \bar{x}_G)$
we show the optimal value $\OPT_\sigma$ is increasing
w.r.t. $\sigma$; 
for $\sigma\in[\bar{x}_G, \sigma_H)$,
we show the optimal value $\OPT_\sigma$ is decreasing
w.r.t. $\sigma$.
The analysis of other case 
where $\sigma_\NI > \bar{x}_G$
follows similarly. 
To show the monotoncity of $\OPT_\sigma$, we 
first characterize optimal solution $F_\sigma$ for 
any $\sigma\in[\sigma_{\NI}, \sigma_H)$, and then 
examine the optimal deviation value $\OPT_\sigma$ under
the deviation $F_\sigma$.
In the remaining of the paper, 
due to the space limit, 
we mainly present the proof for first regime of the 
case $\sigma_\NI \le \bar{x}_G$.

\begin{lemma}
\label{opt_deviation_structure}
Given a prior $H$, and distribution $G$ 
satisfying the 
conditions $(i)$--$(ii)$ in \Cref{thm_nash_iff},
when $\sigma_\NI \le\bar{x}_G$, 
then for any 
$\sigma \in [\sigma_{\NI}, \bar{x}_G]$,
a distribution $F_\sigma$
that satisfies following structure
is an optimal solution to the 
program~\eqref{new_program_general} 
\begin{equation}
\begin{aligned}
    \label{opt_deviation}
    F_\sigma(x) & = 
    \left\{\begin{aligned}
    & H(x), && \forall x \in [0, x^\dagger)\\ 
    & H(x^\dagger) , && \forall x \in [x^\dagger, x^\ddag)\\
    & 1 , && \forall x \in [x^\ddag, 1]\\
    \end{aligned}\right.
\end{aligned}
\end{equation}
where $x^\dagger$ satisfies that 
$\int_0^\sigma F_\sigma(x)dx  
= \sigma - (\lambda - c)$. 
Furthermore, the optimal value $\OPT_\sigma$ 
is increasing w.r.t. $\sigma\in[\sigma_{\NI}, \bar{x}_G]$.
\end{lemma}
The structure of the optimal solution $F_\sigma$ 
admits the following interpretations. 
Let $u(x) := \min\left\{G(x)^{n-1}, G(\sigma)^{n-1}\right\}$.
As we can see, for any $\sigma \le \bar{x}_G$,
$u$ is convex over 
$[0, \sigma]$ 
(recall the convexity $G^{n-1}$ 
in $[0, \bar{x}_G]$)
and is constant over $[\sigma, 1]$. 
Then if a solution $F$ has support 
below $ \sigma$, ideally, 
by Jensen's inequality,
$F$ should allocate its support as much dispersed 
as possible in this interval.
In other words, the MPS constraint should bind
for the support of $F$ that is in 
$[0,  \sigma]$.
At the same time, 
$u$ attains maximum for any values above $\sigma$, 
$F$ thus should put as much mass as possible 
above $ \sigma$. 
Due to the equal-mean constraint 
(i.e., $\int xdF(x) = \lambda$), 
$F$ should put their support that is 
in $[0,  \sigma]$ 
as close to $0$ as possible 
(and simultaneously as much dispersed as possible) 
so that $F$ can allocate more mass above 
$\sigma$. 
Note that the constraint $\sigma_F = \sigma$ is a
linear constraint, and it thus 
determines the cutoff $x^\dagger$
of the portion where $F$ 
satisfies the property in \Cref{lem_sigma_defn}.


For the value $\OPT_\sigma$ for 
$\sigma\in[\bar{x}_G,\sigma_H)$, we show that it
is monotone decreasing w.r.t. 
$\sigma \in[\bar{x}_G,\sigma_H)$.
\begin{restatable}{lemma}{optdeviationpayoff}
\label{opt_deviation_payoff}
For any prior $H$, given a strategy $G$
that satisfies conditions $(i)$--$(ii)$ in \Cref{thm_nash_iff},
the value $\OPT_\sigma$
is monotone decreasing w.r.t. $\sigma \in [\bar{x}_G,\sigma_H)$.
\end{restatable}
To prove this result, for each $\sigma \in[\bar{x}_G,\sigma_H)$,
we first characterize the 
optimal solution $F_\sigma$ to 
the program~\eqref{new_program_general} 
using a much more involved duality argument
(see \Cref{opt_deviation_structure_large_sigma}
and its proof in \Cref{proof-sec-continuous}).
Then with the obtained $F_\sigma$, 
we prove the monotonicity of $\OPT_\sigma$.
The proof  uses the 
convexity of $G^{n-1}$ over $[0, \bar{x}_G]$, 
and is in \Cref{proof-sec-continuous}.
Combine \Cref{opt_deviation_structure} and \Cref{opt_deviation_payoff}
will prove \Cref{lem_most_profitable_devia}. 
Putting all pieces together
can prove \Cref{thm_nash_iff} 
(see the end of \Cref{proof-sec-continuous}).

\section{Applications and Implications of \texorpdfstring{\Cref{thm_nash_iff}}{}}
\label{sec-implications}

\newcommand{\expPara}{\mu}

\newcommand{\gaussianMean}{\lambda}
\newcommand{\gaussianVarPara}{\upsilon}

\newcommand{\LaplaceMean}{\lambda}
\newcommand{\LaplaceVarPara}{b}

In this section, we discuss implications and provide applications of \Cref{thm_nash_iff}.
\subsection{The Effect of the Competition and Inspection Cost on the Equilibrium}
\label{sec-implications}
\Cref{thm_nash_iff} provides a general characterization of the equilibrium for competitive information design for Pandora's Box.
Here we discuss the applications of the theorem
in some interesting/important cases and their implications.
Proofs in this section are in \Cref{proof-sec-implications}.

First of all, as discussed in \Cref{agent_opt_welfare}, 
every sender deploying \essienFull\ is a desired equilibrium as it leads to the highest agent payoff and the highest social welfare. 
Utilizing \Cref{thm_nash_iff}, we can characterize the sufficient and necessary condition for \essienFull\ to be the equilibrium.

\begin{restatable}{corollary}{fullinfornashiif}
\label{full_infor_nash_iif}
Essentially full information strategy is 
the equilibrium strategy if and only if 
$H^{n-1}$ is convex over $[0, \sigma_H]$.
\end{restatable}

\wtr{Intuitively, when all other
senders use the essentially full information strategy, 
sender $i$'s expected payoff by using a strategy $G_i$
can be characterized as follows:
$\sfrac{\sum_{j=0}^{n-1}H(\sigma_H)(1 - H(\sigma_H))^j}{n} 
+ \int_0^{\sigma_H} H(x)^{n-1}dG_i(x)$.
Thus, to maximize the expected payoff, it suffices to 
maximize the expected payoff $\int_0^{\sigma_H} H(x)^{n-1}dG_i(x)$ 
whenever realizing a prize 
$x$ whose  value is smaller than $\sigma_H$.
Now note that when $H^{n-1}$ is convex over $[0, \sigma_H]$,
by Jensen's inequality, sender $i$ would strictly prefer to 
spread the strategy $G_i$ as much as possible 
since it leads to higher payoff. 
Thus, according to \Cref{lem_MPS_singaling}, 
it is optimal for the sender $i$ to 
also use essentially full information strategy.
}

We can also observe a couple of interesting implications of 
\Cref{full_infor_nash_iif}. 
First, increasing competition makes it more likely to reach essential full information disclosure.
This implication is from the the fact that when we fix inspection cost, 
the shape of the function $H^{n-1}$ becomes more convex 
as $n$ increases. 
Moreover, for an arbitrary prior $H$ and 
any cost, one can show that there always 
exists a number of senders such that 
essentially full information is the equilibrium. 
We can also show that for any prior $H$, 
as long as the number of senders is high enough, 
\essienFull\ can be the equilibrium strategy, as formalized below.
\begin{corollary}
\label{large_n_convex}
For any prior $H$ and cost $c \ge 0$,
there exists a $\underline{n} \in \N_+$, 
such that for any $n\ge \underline{n}$, 
\essienFull\ is the equilibrium strategy. 
\end{corollary}
Another implication of \Cref{full_infor_nash_iif} is that, increasing inspection cost makes it more likely to reach essential full information disclosure.
This implication follows from
when we fix the number of senders, 
if essentially full information 
is the equilibrium with a smaller inspection cost,
it is also the equilibrium with a larger cost. 
This is because when increasing the cost, the 
corresponding reservation value $\sigma_H$ 
is decreasing. Therefore, if $H^{n-1}$ 
is already convex on a larger interval $[0, \sigma_H]$,
it  is also
convex on a smaller interval.
To illustrate this observation, 
for a general class of priors -- the prior that has single-peaked density -- 
we can characterize the lower bound cost for the essentially full information to be the equilibrium. 
In particular, when $H^{n-1}$ has single-peaked density,\footnote{
As long as the density function $h$ is log-concave
over $[0, \sigma_H]$, $H^{n-1}$
has single-peaked density over $[0, \sigma_H]$ for any $n$.}
it is always first convex and then concave 
(see example in \Cref{Fig: ex_no_Nash_b}). 
Thus, as long as the reservation value $\sigma_H$ falls
below the inflection point (i.e., the point where
the function $H^{n-1}$ changes from being convex to concave) 
of $H^{n-1}$, 
essentially full information is the equilibrium. 
\begin{corollary}
\label{signle-peaked-full-infor-nash}
Fix $n$ and $H$ such that $H^{n-1}$ 
has single-peaked density over $[0, \sigma_H]$ and its inflection point $\underline{x}$,
let $\underline{c}$ be 
an inspection cost where 
$\sigma_H = \underline{x}$,
then for 
any cost $c \ge \underline{c}$,
essentially full information is the equilibrium. 
\end{corollary}

\wtr{
In below, we exemplify the use of \Cref{signle-peaked-full-infor-nash}
to identify the condition 
of the inspection cost 
for common distributions that admit the existence of 
essentially full information equilibrium strategy
when there are two senders.
\begin{example}[Uniform Prior]
Suppose $H$ is the uniform prior over $[a, b]$ with $a \ge 0$, 
it can be shown that for any inspection cost $c \ge 0$, 
essentially full information strategy is an equilibrium strategy, namely,
a strategy $G = H$ satisfies all conditions in \Cref{thm_nash_iff}.
\end{example}

\begin{example}
[Gaussian Prior]
\label{gaussian_prior_ex}
Suppose $H$ is the Gaussian prior with mean $\gaussianMean > 0$ and variance $\gaussianVarPara^2$ where $\gaussianVarPara \ge 0$, 
it can be shown that
essentially full information strategy is an equilibrium strategy 
if and only if the inspection cost $c$ satisfies
$c \ge \sfrac{\gaussianMean}{2} + \sfrac{\gaussianVarPara}{\sqrt{2\pi}}$.
\end{example}

\begin{example}
[Laplace Prior]
\label{laplace_prior_ex}
Suppose $H$ is the Laplace prior with mean $\gaussianMean > 0$ and 
scale parameter $\LaplaceVarPara > 0$, 
it can be shown that
essentially full information strategy is an equilibrium strategy 
if and only if the inspection cost $c$ satisfies
$c \ge \sfrac{\LaplaceVarPara}{2}$.
\end{example}
Intuitively, fix an inspection cost $c > 0$ and the prior mean, 
both \Cref{gaussian_prior_ex} and \Cref{laplace_prior_ex}
suggest that it is more likely to have 
essentially full information strategy as the equilibrium strategy
when the prior distribution has smaller variance. 
}

In addition to characterizing the equilibrium conditions, we can also show that, under the condition that
essentially full information is the equilibrium, the agent’s payoff decreases as the inspection cost increases and
increases as the number of senders increases.
\begin{restatable}{corollary}{essentialfullpayoff}
\label{essential_full_payoff}
Under essentially full information equilibrium, 
the agent's payoff is decreasing with respect to the inspection cost, 
and increasing with respect to the number of senders.
\end{restatable}
\wtr{
Intuitively, the above results follow from the 
fact that agent's expected payoff  
$\sigma_H - \int_0^{\sigma_H}H(x)^ndx$
under essentially full information equilibrium
is an increasing function over the reservation value $\sigma_H$,
which is decreasing with respect to the inspection cost;
and the payoff is an increasing function 
with respect to the number of senders.
}

Below we provide one more example on how \Cref{thm_nash_iff} can help us characterize the equilibrium in different cases.
When $H^{n-1}$ is concave 
over $[0, \sigma_H]$, using the conditions
$(i)$--$(ii)$, we can characterize a unique distribution 
$G$ such that $G^{n-1}$ will be firstly linear
over $[0, \bar{x}_G]$
and then flat over $[\bar{x}_G, \sigma_H]$ 
(see the example in \Cref{Fig: ex_no_Nash_a}). 
Using the linearity of $G^{n-1}$,
we can show that to verify whether such 
$G$ is an equilibrium strategy, it only suffices 
to check whether $G(\lambda -c)^{n-1}\le \sfrac{1}{n}$.

\begin{restatable}{corollary}{concaveNashiif}
\label{concave_Nash_iif}
Given prior $H$ such that 
$H^{n-1}$ is concave over $[0, \sigma_H]$. 
Let $G$ be a distribution 
satisfying the conditions 
$(i)$--$(ii)$ in \Cref{thm_nash_iff}, 
then $G$ is an equilibrium strategy if and only if 
$G(\lambda - c)^{n-1} \le \sfrac{1}{n}$.
\end{restatable}
\wtr{
We also exemplify below the use of \cref{concave_Nash_iif}
to identify the condition 
of the inspection cost 
for common distribution that admit the existence of 
equilibrium strategy
when there are two senders.
\begin{example}
[Exponential Prior]
\label{ex:exp prior}
Suppose $H$ is the exponential prior over $[0, +\infty)$ with the parameter 
$\expPara \ge 0$, namely, $H(x) = 1 - \exp(-\expPara x)$.
Since $H$ is concave over the whole support $[0, +\infty)$, 
it can be shown that there exists an equilibrium strategy 
$G$ (in particular, one can deduce the behavior of strategy $G$ 
over $[0, \sigma_H]$ where 
$\sigma_H = \sfrac{-\ln (\expPara c)}{\expPara}$ 
as follows: 
$G(x) = \frac{H(\sigma_H)}{2\sigma_H - \frac{2(\sigma_H - (\sfrac{1}{\expPara} - c))}{H(\sigma_H)}} x, \forall x\in [0, 2\sigma_H - \frac{2(\sigma_H - (\sfrac{1}{\expPara} - c))}{H(\sigma_H)}]; 
G(x) = H(\sigma_H), \forall x \in  [2\sigma_H - \frac{2(\sigma_H - (\sfrac{1}{\expPara} - c))}{H(\sigma_H)}, \sigma_H]$) 
if and only if 
the inspection cost $c \ge 0$ and the parameter $\expPara$
satisfy $(\expPara c)^3 - 3 (\expPara c)^2 + 
2\expPara c + \expPara c\ln (\expPara c) \ge 0$.
Note that when fixing any inspection cost $c > 0$, 
function $(x c)^3 - 3 (x c)^2 + 
2x c + x c\ln (x c)$ crosses $x$-axis over $(0, \sfrac{1}{c})$
once and it crosses from below.
Intuitively, this suggests that for any fixed cost $c > 0$, 
it is more likely to admit the existence of a
symmetric equilibrium if the parameter $\expPara$ is larger, 
i.e., the  prior has smaller variance.
\end{example}
}

\subsection{Informational Properties of Pandora's Box}
\label{sec-prelim}

In this section, we investigate how senders' 
\infoStrategies\ affect the agent's payoff
under optimal inspection strategy and 
how agent's optimal payoff can be used to inform the informativeness of box's information
strategy. 
To this end, we provide several properties 
about the reservation values
which will be useful for our later equilibrium analysis
in \Cref{sec-continuous}. 
While reservation values have been
well-studied in the Pandora's Box problem, 
to our knowledge, the informational properties we present in this section are not known before. 

Below we present the main result in this section,
which characterizes an intrinsic connection between 
informativeness of any box's value distribution and 
the optimal payoff order of the search agent.
We use $\agentU(G_i, G_{-i})$ to denote agent's optimal expected payoff 
under the boxes' strategies $(G_i, G_{-i})$ where 
$G_{-i} := (G_j)_{j\in[n], j\neq i}$ 
contains all boxes' prize distributions excluding box $i$'s prize distribution.
\begin{theorem}
\label{thm_r_utility_sigma}
The distribution $G_i'$ is a mean-preserving spread of distribution $G_i$, 
i.e., $G_i' \blackwellOrder G_i$, if and only if 
$\agentU(G_i', G_{-i}) \ge \agentU(G_i, G_{-i})$ 
for all other boxes' strategies $G_{-i}$, all $(c_i)_{i\in[n]}$,
and $G_i, G_i'$ have the same mean. 
\end{theorem}
It is worth noting that 
the above results do not require any assumption of symmetric prior prize distribution or symmetric cost for opening all boxes.
Intuitively, the ``if'' part of the above results provides 
another way to compare the Blackwell's informativeness via comparing the 
agent's optimal payoff in a basic setup of Pandora's Box problem
(recall that the Blackwell's informativeness says that 
a \infoStrategy\ $G'$ is \emph{more informative} 
than $G$ if $G'$ is an MPS of $G$).
The ``only if'' part of the above results shows that 
the agent obtains a higher payoff whenever a box becomes more informative.
With this implication, an important corollary is that, 
when all boxes are performing \fullInforStrategy, 
i.e., $G_i = H_i$ for all $i$, 
the agent obtains the highest payoff. 
Below we demonstrate a stronger version of this claim. 
In particular, we define the following \emph{essentially} full information strategy 
which fully reveals information whenever the prize
value is no larger than the reservation value of 
this information strategy:
\begin{definition}[Essentially Full Information Strategy]
For any box $i\in[n]$,
a strategy $G: H_i \blackwellOrder G$ is \essienFull\ for box $i$
if $G$ satisfies that $G(x) = H_i(x), \forall x\in[0, \sigma_{H_i}]$,
where $\sigma_{H_i}$ is the reservation value of the prior $H_i$.
\end{definition} 

We can show that, for the agent to achieve the highest payoff, it suffices that all senders use \essienFull.
\begin{corollary}
\label{agent_opt_welfare}
Let $G_i$ be an \essienFull\ for box $i\in[n]$.
Then agent obtains the highest expected payoff 
under $(G_1, \ldots, G_n)$ among all possible 
(symmetric or asymmetric) strategy profiles.
Moreover, when $H \equiv H_i, \forall i\in[n]$, 
the agent's highest expected payoff can be computed as 
$\sigma_H - \int_0^{\sigma_H}H(x)^ndx$.
\end{corollary}
The basic intuition behind the above \Cref{agent_opt_welfare}
is that in Pandora's Box,
when the agent uses the 
optimal inspection strategy,
after she inspects sender $i$, 
as long as the mean of the posterior 
for sender $i$ after inspection is higher than his reservation value,
the agent will take the same action: 
stop inspection and choose sender $i$. 
This observation implies that 
the distribution above the reservation value 
of the sender's \infoStrategy\
does not change the agent's decisions and payoffs.

Note that since the agent chooses exactly one sender at the end,
the total payoff to all senders is $1$ 
no matter what the agent's inspection strategy is 
and what the senders' \infoStrategies\ are. 
Therefore, when all senders use \essienFull, it not only 
maximizes the agent's payoff, it also achieves
the maximum social welfare.
Given this desired property for \essienFull, 
in \Cref{sec-continuous}, we characterize the sufficient 
and necessary condition for all senders to use \essienFull\
(see \Cref{full_infor_nash_iif}) in equilibrium. 

\xhdr{Additional useful properties}
Before presenting the proof of \Cref{thm_r_utility_sigma}, 
we describe a few other informational properties of Pandora's Box. 
First, recall that
we say a distribution $G'$ is \emph{more informative} 
than $G$ if $G'$ is an MPS of $G$, i.e., $G' \blackwellOrder G$.
This partial order of informativeness is from Blackwell's information theorem~\citep{Blackwell-53}.
En route to proving \Cref{thm_r_utility_sigma}, 
we also show the following total order on the reservation values induced by information strategies.

\begin{proposition}
\label{prop_informativness_sigma}
For any cost $c\ge 0$ and two distributions $G'$ and $G$, if $G'\blackwellOrder G$, $\sigma_{G'} \ge \sigma_G$.
\end{proposition}
That is,  a more informative   sender  \infoStrategy\ 
leads to a higher reservation value. Since the agent inspects the senders in an decreasing order of their reservation values, the proposition confirms the intuition 
that the agent would first inspect the sender who uses more informative \infoStrategy.
Below we give the lower and upper bounds of the reservation values for any feasible sender's \infoStrategy\ $G$ given prior of this sender. 
Moreover, we provide conditions on when 
the sender's \infoStrategy\ $G$ has the
lowest or highest reservation value, corresponding to the most uninformative or most informative \infoStrategy.
\begin{restatable}{corollary}{intervalMPS}
\label{interval_MPS}
Fix any box $i$,
given the prior $H_i$ and the cost $c_i \ge 0$, 
for any \infoStrategy\ $G$ that
satisfies $H_i \blackwellOrder G$, we have 
$\lambda_i - c_i \leq \sigma_G \leq \sigma_{H_i}$
where $\lambda_i = \expect[x\sim H_i]{x}$. Moreover, 
\begin{itemize}
    \item $\sigma_G=\lambda_i-c$ if and only if $G$ has no support over $[0,\lambda_i - c_i]$;
    \item $\sigma_G=\sigma_{H_i}$ if and only if $H_i$ is an MPS of $G$ over the interval\footnote{Let $W(y) := \int_0^y \big[H_i(x) - G(x)\big] dx$. We say $H_i$ is an MPS of $G$ over $[a, b]$ if and only if $W(a) = W(b) = 0$, and $W(y) \ge 0, \forall y \in [a, b]$. } $[0, \sigma_{H_i}]$,  denoted by $H_i\blackwellOrder_{[0, \sigma_{H_i}]} G$.
\end{itemize}
\end{restatable}
The above corollary characterizes the sender's \infoStrategies\ 
that reach the lowest and highest reservation values. 
We should expect when the sender uses 
no (full) information \infoStrategy, 
the strategy should lead to the lowest (highest) reservation value. 
As a sanity check,
when the sender $i$ uses \noInforStrategy, 
the corresponding $G$ contains a single point mass at $\lambda_i$,
and it is easy to see that corresponding reservation value is $\lambda_i - c_i$.
When the sender uses \fullInforStrategy, 
i.e., the corresponding $G$ equals to the prior, 
the reservation value is $\sigma_{H_i}$. 

We provide a proof overview of \Cref{thm_r_utility_sigma}.
In the agent's optimal inspection strategy
(as specified in \Cref{thm_optimal_searching}), 
both the selection rule and the stopping rule 
depend on the reservation value. 
To see how the agent's payoff changes
if one sender uses a different \infoStrategy,
one needs to understand how the reservation
value ties with sender's \infoStrategy. 
Thanks to \Cref{thm_r_utility_sigma}, we know that the reservation value is always weakly larger if the \infoStrategy\
is more informative (see \Cref{prop_informativness_sigma}). 
With this result, armed with an 
already known result which shows
the expected payoff of any   inspection policy is 
bounded above by the expectation of 
highest ``capped'' reservation value 
(see \Cref{agent_opt}), 
we can then prove \Cref{thm_r_utility_sigma}.

We conclude this section by noting that our proof for the ``only if'' direction essentially shows that the capped value  of a more informative strategy is \emph{second-order} stochastically dominated by the capped value of a less informative strategy.
Then by the convexity of the maximum operator, 
one can also achieve the ``only if'' result of \Cref{thm_r_utility_sigma}.

\section{Conclusion}
\label{sec-conclusion}


In this paper, we study the competitive information design
for the Pandora's Box problem. 
We characterize the informational properties
of Pandora's Box by analyzing 
how a box's partial information disclosure 
affects the agent's optimal decisions.
We fully characterize the pure symmetric equilibrium 
for the boxes' competitive information disclosure 
with providing necessary and sufficient conditions that guarantee the existence and uniqueness of competition equilibrium, and reveal various insights regarding information competition and the resultant agent payoff at equilibrium.

\bibliography{mybib}

\newpage
\appendix

\newcommand{\opponentSrategy}{G}
\newcommand{\deviationSrategy}{F}

\section{Missing proof of \Cref{sec-opt-searching}}
\label{proof-sec-opt-searching}

\newcommand{\realizedVal}{\tilde{x}}
\begin{proof}[Proof of \Cref{thm:agent opt}]
For each box $i\in[n]$, let $\{\messageprob_i(\cdot \condition x), \messagespace_i\}$ be the sender $i$'s signaling mechanism.
Given a signal realization $q_i \sim \messageprob_i(\cdot \condition x)$, together with the prior prize distribution $\priorCDF_i$, the agent will form a posterior belief, denoted by $\mu_i(\cdot \mid q_i) \in \Delta([0, 1])$, for the underlying realized prize. 
Let $x_i \triangleq \expect[x\sim \mu_i(\cdot \mid q_i)]{x}$ be the corresponding posterior mean.
Slightly abusing the notation, we use $\messageprob_i(\cdot) \in \Delta(\messagespace_i)$ to denote the marginal distribution of realized signals. 
Let $G_i\in \Delta([0, 1])$ be the corresponding distribution of posterior mean jointly induced from the $\{\messageprob_i(\cdot \condition x), \messagespace_i\}$ and the prior prize distribution. 
Let $\realizedVal_i\in[0, 1]$ be the realized prize value if the agent decides to take the prize from box $i$.
Clearly, by definition, we have $x_i\sim G_i$.
Let $\pi$ be the agent's any search and stopping strategy, then we can express the agent's expected payoff from the strategy $\pi$ as follows:
\begin{align*}
    & \expect[(\selectIdx_i, \openIdx_i) \sim \pi, \forall i ]{\sum_{i\in[n]} \expect[\realizedVal_i \sim \priorCDF_i]{\selectIdx_i \realizedVal_i - \openIdx_i c_i}} \\
    = ~ & 
    \expect[(\selectIdx_i, \openIdx_i) \sim \pi, \forall i ]{\expect[q_i \sim \messageprob_i(\cdot), \forall i]{\sum_{i\in[n]} \expect[\realizedVal_i \sim \mu_i(\cdot\mid q_i)]{\selectIdx_i \realizedVal_i - \openIdx_i c_i}}} \\
    = ~ & 
    \expect[(\selectIdx_i, \openIdx_i) \sim \pi, \forall i ]{\expect[q_i \sim \messageprob_i(\cdot), \forall i]{\sum_{i\in[n]} \selectIdx_i \expect[\realizedVal_i \sim \mu_i(\cdot\mid q_i) ]{\realizedVal_i} - \openIdx_i c_i}}\\
    = ~ & 
    \expect[(\selectIdx_i, \openIdx_i) \sim \pi, \forall i ]{\expect[x_i\sim G_i, \forall i]{\sum_{i\in[n]} \selectIdx_i x_i - \openIdx_i c_i}} \\
    = ~ & 
    \sum_{i\in[n]} \expect[x_i\sim G_i]{\expect[(\selectIdx_i, \openIdx_i) \sim \pi]{\selectIdx_i x_i - \openIdx_i c_i}}
\end{align*}
We next proceed the proof by establishing the following upper bound
\begin{align*}
    \sum_{i\in[n]} \expect[x_i\sim G_i]{\expect[(\selectIdx_i, \openIdx_i) \sim \pi]{\selectIdx_i x_i - \openIdx_i c_i}} 
    \le 
    \sum_{i\in[n]} \expect{\selectIdx_i \kappa_i}
\end{align*}
where $\kappa_i$ for every $i\in[n]$ satisfies that 
$\kappa_i \triangleq \min\{x_i, \sigma_{G_i}\}$ and the expectation in right hand side of above inequality is over all randomness.
To see this, let $b_i \triangleq \max\{x_i  - \sigma_{G_i}, 0\}$.
Then by definition, we have $\expect{b_i} = c_i$, and also we have
\begin{align*}
    \sum_{i\in[n]} \expect[x_i\sim G_i]{\expect[(\selectIdx_i, \openIdx_i) \sim \pi]{\selectIdx_i x_i - \openIdx_i c_i}} 
    & = 
    \sum_{i\in[n]} \expect[x_i\sim G_i]{\expect[(\selectIdx_i, \openIdx_i) \sim \pi]{\selectIdx_i (\kappa_i + b_i) - \openIdx_i \expect{b_i}}}\\
    & = \sum_{i\in[n]} \expect[x_i\sim G_i]{\expect[(\selectIdx_i, \openIdx_i) \sim \pi]{\selectIdx_i (\kappa_i + b_i) - \openIdx_i b_i}}\\
    & =
    \sum_{i\in[n]} \expect[x_i\sim G_i]{\expect[(\selectIdx_i, \openIdx_i) \sim \pi]{\selectIdx_i \kappa_i - (\openIdx_i - \selectIdx_i) b_i }}\\
    & \le 
    \sum_{i\in[n]} \expect[x_i\sim G_i]{\expect[(\selectIdx_i, \openIdx_i) \sim \pi]{\selectIdx_i \kappa_i}}~.
\end{align*}
Here in second equality, we observe that the decision variable $\openIdx_i$ on whether to inspect box $i$ can only depend on information that is not dependent on the realized posterior mean $x_i$. 
Thus, we have $\expect{\openIdx_i \expect{b_i}} = \expect{\openIdx_ib_i}$.
The last inequality is due to the fact that $b_i \ge 0, \selectIdx_i \le \openIdx_i$ for all $i$.
Since $\sum_i \selectIdx_i \le 1$, for any agent's strategy, we further have the following upper bound of agent's expected payoff: 
\begin{align*}
     \sum_{i\in[n]} \expect[x_i\sim G_i]{\expect[(\selectIdx_i, \openIdx_i) \sim \pi]{\selectIdx_i \kappa_i}} 
     \le 
     \expect[x_i \sim G_i, \forall i]{\max_{i\in[n]} \kappa_i}~.
\end{align*}
We next argue that the strategy characterized in \Cref{thm:agent opt} can indeed obtain above expected payoff upper bound $\expect{\max_i \kappa_i}$. 
To see this, consider that the strategy characterized in \Cref{thm:agent opt} ends up with selecting box $i$ to take the prize value, but instead there exists another box $j$ that has $\kappa_j > \kappa_i$. We below show that this case cannot happen.
There are two possible cases
\begin{enumerate}
    \item[(i)] $\sigma_{G_i} < \sigma_{G_j}$: in this case, the agent first inspects box $j$. 
    Consider following two scenarios: 
    (a) If the realized posterior mean $x_j \ge \sigma_{G_j}$, then the agent would have to select box $j$ as $x_j$ is larger than all remaining reservation values of uninspected boxes.  
    (b) If the realized posterior mean $x_j < \sigma_{G_j}$, then $\kappa_j = x_j$, and we have $x_j > \kappa_i$ by assumption. Meanwhile, we must also have $x_j < \sigma_{G_i}$ as the agent would never inspect box $i$. However,  since we also have $x_j > \kappa_i$, together with $x_j < \sigma_{G_i}$, we also have $x_j > x_i$, which also leads to a contradiction as the agent would have to select box $j$ to take the prize. 
    \item[(ii)] $\sigma_{G_i} \ge \sigma_{G_j}$: in this case, the agent first inspects box $i$. 
    Consider following two scenarios: 
    (a) If the realized posterior mean $x_j \ge \sigma_{G_j}$, then we have $\kappa_j = \sigma_{G_j} > \kappa_i$.
    When $x_i \le \sigma_{G_i}$, then agent would not select box $i$ to take the prize as $x_i =\kappa_i < \sigma_{G_j}$.
    When $x_i > \sigma_{G_i}$, then $\kappa_i = \sigma_{G_i} < \sigma_{G_j}$ which leads to a contradiction. 
    (b) If the realized posterior mean $x_j < \sigma_{G_j}$, 
    then we have $\kappa_j = x_j > \kappa_i$.
    Similarly, when $x_i \le \sigma_{G_i}$, then agent would not select box $i$ to take the prize as $x_i =\kappa_i < x_j < \sigma_{G_j}$.
    When $x_i > \sigma_{G_i}$, then $\kappa_i = \sigma_{G_i} < x_j < \sigma_{G_j}$ which leads to a contradiction. 
\end{enumerate}
Putting above pieces together, we can show that the agent strategy described in \Cref{thm:agent opt} can indeed achieve highest payoff upper bound $\expect[x_i \sim G_i, \forall i]{\max_{i\in[n]} \kappa_i}$. We thus finish the proof.
\end{proof}


\section{Missing Proofs for \Cref{sec-continuous}}
\label{proof-sec-continuous}

\begin{proof}[Proof of \Cref{max_sigma_nash}]
We prove the lemma using two senders case. 
The analysis for multiple senders 
can be easily carried over.
Given a symmetric strategy $(\opponentSrategy, \opponentSrategy)$ where $\sigma_\opponentSrategy \neq \sigma_H$,
let $\bar{x}_\opponentSrategy = \max\{x: x\in\supp{\opponentSrategy} \wedge x\le\sigma_\opponentSrategy\}$, 
we now consider following possible scenarios:
\squishlist
    \item $\opponentSrategy(x) = H(x), \forall x\in[0, \bar{x}_\opponentSrategy]$. In this case, we must have $\bar{x}_\opponentSrategy < \sigma_\opponentSrategy$, otherwise we have $\sigma_\opponentSrategy = \sigma_H$.
    Consider (sufficiently small) $\varepsilon$ and $\varepsilon'$, 
    and let $x^\dagger := \min\{x: \opponentSrategy(x) \ge H(\bar{x}_\opponentSrategy + \varepsilon)\}$.
    Consider sender $1$ deviating to a new strategy $\deviationSrategy$ where 
    \begin{equation*}
    \begin{aligned}
        \deviationSrategy(x) & = 
        \left\{\begin{aligned}
        & \opponentSrategy(x), && \forall x \in [0, \bar{x}_\opponentSrategy)\\ 
        & H(x) , && \forall x \in [\bar{x}_\opponentSrategy, \bar{x}_\opponentSrategy + \varepsilon)\\
        & H(\bar{x}_\opponentSrategy + \varepsilon) , && \forall x \in [\bar{x}_\opponentSrategy + \varepsilon,x^\dagger + \varepsilon')\\
        & \opponentSrategy(x), && \forall x \in [x^\dagger + \varepsilon', 1],
        \end{aligned}\right.
    \end{aligned}
    \end{equation*}
    where $\varepsilon'$ further satisfies that
    \begin{align*}
        \int_{\bar{x}_\opponentSrategy}^{x^\dagger} (\deviationSrategy(x) - \opponentSrategy(x)) dx = 
        \int_{x^\dagger}^{x^\dagger + \varepsilon'} (\opponentSrategy(x) - \deviationSrategy(x)) dx~.
    \end{align*}
    By construction, we have $\deviationSrategy \blackwellOrder \opponentSrategy$ 
    as $\int_0^\sigma (\deviationSrategy(x) - \opponentSrategy(x)) dx \ge 0, 
    \forall \sigma$,
    and $H\blackwellOrder\deviationSrategy$ as 
    $\int_0^\sigma (H(x) - \deviationSrategy(x)) dx \ge 0, \forall \sigma$.
    Let $\Delta_\varepsilon := H(\bar{x}_\opponentSrategy + \varepsilon) - H(\bar{x}_\opponentSrategy)$.
    Now consider 
    \begin{align*}
        \int_{\sigma_\opponentSrategy}^1(x - \sigma_\opponentSrategy) d\deviationSrategy(x) - \int_{\sigma_\opponentSrategy}^1(x - \sigma_\opponentSrategy) d\opponentSrategy(x) 
        & =   \int_{0}^{\sigma_\opponentSrategy} \deviationSrategy(x)dx - \int_{0}^{\sigma_\opponentSrategy} \opponentSrategy(x)dx > 0, \\
        \Rightarrow ~~ 
        \int_{\sigma_\opponentSrategy}^1(x - \sigma_\opponentSrategy) d\deviationSrategy(x)  & > c~.
    \end{align*}
    As $\int_\sigma^1 (x - \sigma)d\deviationSrategy(x)$ is strictly decreasing w.r.t $\sigma$, 
    we thus have $\sigma_\deviationSrategy > \sigma_\opponentSrategy$. 
    Now let $\senderU_a := p_\opponentSrategy + \int_0^{\sigma_\opponentSrategy} \opponentSrategy(x) d\opponentSrategy(x)$ and consider
    \begin{align*}
        \senderU_1(\deviationSrategy, \opponentSrategy) - u_a^S
        & = \int_{\sigma_\opponentSrategy}^1 d\deviationSrategy(x) + \int_0^{\sigma_\opponentSrategy} \opponentSrategy(x)d\deviationSrategy(x) - u_a^S\\
        & = \int_{\bar{x}_\opponentSrategy}^{\bar{x}_\opponentSrategy + \varepsilon} \opponentSrategy(x)d\deviationSrategy(x)
        - \Delta_\varepsilon = (1 - p_\opponentSrategy)
        - \Delta_\varepsilon = - p_\opponentSrategy \Delta_\varepsilon 
    \end{align*}
    Choose $\varepsilon$ such that $p_\opponentSrategy \Delta_\varepsilon < u_a^S -  \frac{1}{2}$, we then have
    \begin{align*}
        \senderU_1(\deviationSrategy, \opponentSrategy) = u_a^S - p_\opponentSrategy \Delta_\varepsilon > \frac{1}{2} = \senderU_1(\opponentSrategy, \opponentSrategy).
    \end{align*}
    
    \item $\exists x\in[0, \bar{x}_\opponentSrategy], \opponentSrategy(x) \neq H(x)$. 
    In this case, we consider two possible scenarios:
    \begin{enumerate}
        \item When $\opponentSrategy(\sigma_\opponentSrategy) > H(\sigma_\opponentSrategy)$. In this case, as we have 
        $\int_0^{\sigma_\opponentSrategy} \opponentSrategy(x)dx < \int_0^{\sigma_\opponentSrategy} H(x)dx$, there must exist 
        a point $x^\dagger := \max\{x \in [0, \bar{x}_\opponentSrategy]: \opponentSrategy(x) \ge H(x) \wedge \opponentSrategy(x) < H(x)\}$. Now consider following new strategy $\deviationSrategy$:
        \begin{equation*}
        \begin{aligned}
            \deviationSrategy(x) & = 
            \left\{\begin{aligned}
            & \opponentSrategy(x), && \forall x \in [0, x^\dagger - \varepsilon)\\ 
            & H(x) , && \forall x \in [x^\dagger - \varepsilon, x^\dagger)\\
            & \opponentSrategy(x) , && \forall x \in [x^\dagger,  x^\ddag)\\
            & \opponentSrategy(x^\ddag) , && \forall x \in [x^\ddag,  x^\ddag + \varepsilon')\\
            & \opponentSrategy(x), && \forall x \in [x^\ddag + \varepsilon', 1],
            \end{aligned}\right.
        \end{aligned}
        \end{equation*}
        where $x^\ddag \ge \sigma_\opponentSrategy$ and $\varepsilon, \varepsilon'$
        are sufficiently small such that they satisfy the following
        \begin{align*}
            \int_{x^\dagger-\varepsilon}^{x^\dagger} (\deviationSrategy(x) - \opponentSrategy(x)) dx = 
            \int_{x^\ddag}^{x^\ddag + \varepsilon'} (\opponentSrategy(x) - \deviationSrategy(x)) dx.
        \end{align*}
        By construction, $\deviationSrategy\blackwellOrder\opponentSrategy$ 
        as $\int_0^\sigma (\deviationSrategy(x) - \opponentSrategy(x)) dx \ge 0, \forall \sigma$,
        and $H\blackwellOrder\deviationSrategy$ as $\int_0^\sigma (H(x) - \deviationSrategy(x)) dx \ge 0, \forall \sigma$.
        Now consider 
        \begin{align*}
            \int_{\sigma_\opponentSrategy}^1(x - \sigma_\opponentSrategy) d\deviationSrategy(x) - \int_{\sigma_\opponentSrategy}^1(x - \sigma_\opponentSrategy) d\opponentSrategy(x)
            & = \int_{0}^{\sigma_\opponentSrategy} \deviationSrategy(x)dx - \int_{0}^{\sigma_\opponentSrategy} \opponentSrategy(x)dx > 0.
        \end{align*}
        Thus, we have $\sigma_\deviationSrategy > \sigma_\opponentSrategy$. 
        As a result, let $\senderU_a := p_\opponentSrategy + \int_0^{\sigma_\opponentSrategy} \opponentSrategy(x) d\opponentSrategy(x)$ and 
        \begin{align*}
            \senderU_1(\deviationSrategy, \opponentSrategy) - u_a^S
            & = \int_{\sigma_\opponentSrategy}^1 d\deviationSrategy(x) + \int_0^{\sigma_\opponentSrategy} \opponentSrategy(x)d\deviationSrategy(x) - u_a^S \\
            & = \int_{x^\dagger - \varepsilon}^{x^\dagger}\opponentSrategy(x) \cdot (h(x) - f(x))dx
        \end{align*}
        Choose $\varepsilon$ such that $\int_{x^\dagger - \varepsilon}^{x^\dagger}\opponentSrategy(x) \cdot (h(x) - f(x))dx < u_a^S -  \frac{1}{2}$, we then have
        \begin{align*}
            \senderU_1(\deviationSrategy, \opponentSrategy) = u_a^S - \int_{x^\dagger - \varepsilon}^{x^\dagger}\opponentSrategy(x) \cdot (h(x) - f(x))dx > \frac{1}{2} = \senderU_1(\opponentSrategy, \opponentSrategy).
        \end{align*}
        \item When $\opponentSrategy(\sigma_\opponentSrategy) \le H(\sigma_\opponentSrategy)$.
        In this case, consider the point $x^\dagger := \max\{x \in [0, \bar{x}_\opponentSrategy]: \opponentSrategy(x) \le H(x) \wedge \opponentSrategy(x) > H(x)\}$.
        Now consider following new strategy $\deviationSrategy$:
        \begin{equation*}
        \begin{aligned}
            \deviationSrategy(x) & = 
            \left\{\begin{aligned}
            & \opponentSrategy(x), && \forall x \in [0, x^\dagger)\\ 
            & H(x) , && \forall x \in [x^\dagger , x^\dagger + \varepsilon)\\
            & H(x^\dagger + \varepsilon) , && \forall x \in [x^\dagger + \varepsilon,  \bar{x})\\
            & \opponentSrategy(x) , && \forall x \in [\bar{x},  x^\ddag)\\
            & \opponentSrategy(x^\ddag) , && \forall x \in [x^\ddag,  x^\ddag + \varepsilon')\\
            & \opponentSrategy(x), && \forall x \in [x^\ddag + \varepsilon', 1],
            \end{aligned}\right.
        \end{aligned}
        \end{equation*}
    \end{enumerate}
    where $x^\ddag \ge \sigma_\opponentSrategy$, and $\bar{x}$ satisfies $\opponentSrategy(\bar{x}) = H(x^\dagger + \varepsilon)$.
    Moreover, $\varepsilon, \varepsilon'$ are sufficiently small such that they satisfy the following
    \begin{align*}
        \int_{x^\dagger}^{\bar{x}} (\deviationSrategy(x) - \opponentSrategy(x)) dx = 
        \int_{x^\ddag}^{x^\ddag + \varepsilon'} (\opponentSrategy(x) - \deviationSrategy(x)) dx.
    \end{align*}
    Follow the earlier analysis, we have $\sigma_\deviationSrategy > \sigma_\opponentSrategy$, 
    and with sufficiently small $\varepsilon, \varepsilon'$, we have $u_1^S(\deviationSrategy, \opponentSrategy) > u_1^S(\opponentSrategy, \opponentSrategy)$.
\squishend
Putting pieces together, the proof then completes.
\end{proof}

\begin{proof}[Proof of \Cref{symme_deviation}]
We first prove the first part of the statement.
Given a prior $H$, 
let $G$ be the distribution satisfying
conditions $(i)$--$(ii)$ in \Cref{thm_nash_iff}.
We now consider sender $i$'s 
best response strategy $F$ that is subject to 
$\sigma_F = \sigma_H$ given 
all other senders using strategy $G$.
For notation simplicity, 
define following quantile value 
$p_F := 1 - F(\sigma_F),
p_G := 1- G(\sigma_G)$, 
and $p_H := 1 - H(\sigma_H)$.
Observe that whenever sender $i$ is inspected, 
there are two possible cases, 
either the realized $x_i \ge \sigma_H$ 
where the agent will stop the inspection 
and claim $x_i$ from sender $i$;
or the realized $x_i < \sigma_H$ where the agent claims 
$x_i$ from sender $i$ only if he inspects all senders 
and finds out $i = \argmax_j x_j$.
With the above observation, 
we have following sender $i$'s expected payoff
on deviating to \infoStrategy\ $F$:
\begin{align*}
    \senderU_i(G, \ldots, F, \ldots, G) 
    & =\sum_{j=1}^n \left(p_F \cdot (1 - p_G)^{j - 1} + \int_0^{\sigma_H} G(x)^{n-1}dF(x)\right)  \cdot \frac{1}{n} \\
    & \labelrel{=}{helper_1_SP} \frac{1}{n} \cdot \sum_{j=0}^{n-1}p_F\cdot(1 - p_G)^j + \int_0^{\sigma_H} G(x)^{n-1}dF(x)~,
\end{align*}
where in~\eqref{helper_1_SP} we use $p_F = p_H = p_G$ 
due to \Cref{lem_same_density}.
Now we consider following sender $i$'s best response problem
that is subject to deviating 
to strategies in $\cH(\sigma_H)$:
\begin{align*}
    \max_{F\in\cH(\sigma_H)} ~
    \frac{1}{n} \cdot \sum_{j=0}^{n-1}p_H\cdot(1 - p_H)^j + \int_0^{\sigma_H} G(x)^{n-1}dF(x)~.
\end{align*}
Given a prior $H$, $p_H$ is a constant. 
The above program can be further 
reduced to
\begin{align}
    \label{program_self_br_general}
    \max_{F\in\cH(\sigma_H)} ~
    \int_0^{\sigma_H} G(x)^{n-1}dF(x)~.
\end{align}
Recall that from \Cref{interval_MPS}, 
the constraint $\sigma_F = \sigma_H$ 
is equivalent to 
requiring $H\blackwellOrder_{[0, \sigma_H]} F$.
To complete the proof, 
we note that \citet{HKB-19} have shown 
when $c = 0$, a \infoStrategy\ $G$ that satisfies 
the properties in \Cref{defn_linear_MPS}
over the interval $[0, 1]$ is the best response
\infoStrategy\ to itself, i.e., 
$G$ is the solution to the program 
$\max_{F\in\cH}\int_0^1 G(x)^{n-1}dF(x)$.
Now given a \infoStrategy\ that satisfies 
the conditions $(i)$--$(iii)$, it is easy to see that
any strategy $G^*$ that satisfies 
$G^*(x) = G(x), \forall x\in[0, \sigma_H]$ is 
the optimal solution to the program~\eqref{program_self_br_general}.
The second part of the statement follows 
from the necessity the equilibrium strategy $G$ 
when $c = 0$ in \citep{HKB-19}.
\end{proof}

\begin{proof}[Proof of \Cref{equilibrium_simplication}]
It suffices to show that given $(G_1, \ldots, G_n)$,
no sender has profitable deviation. 
Consider following two kinds of deviation:
one is deviating to a strategy that has reservation 
value $\sigma_H$, then from \Cref{symme_deviation}, we know
there exists no such profitable deviation;
for any $\sigma < \sigma_H$,
the other is deviating to a strategy that has reservation 
value $\sigma$, then from \eqref{deviation_payoff_smaller_RV}
and the definition of $(G, \ldots, G)$, we know
there exists no such profitable deviation.
\end{proof}

\begin{restatable}{lemma}{optdeviationstructurelargesigma}
\label{opt_deviation_structure_large_sigma}
Given a prior $H$,
and a unique distribution $G$ 
satisfying the 
conditions $(i)$--$(ii)$ in \Cref{thm_nash_iff},
for any $\sigma \in 
[\max\{\sigma_{\NI}, \bar{x}_G\}, \sigma_H)$,
let $\Delta$ satisfy 
$\sigma - (\lambda - c) + H(\sigma+\Delta)\cdot \Delta
= \int_0^{\sigma+\Delta} H(x)dx$,
and let $x^* := x_m$, i.e., the last
point where $G^{n-1}$ is strictly convex,
a distribution $F_\sigma$
satisfying following structure
is an optimal solution to the 
program~\eqref{new_program_general}
\begin{enumerate}
    \item if $\int_0^{x^*} H(x)dx + (\bar{x}_G - x^*) \cdot H(x^*) +
    (\sigma - \bar{x}_G) \cdot H(\sigma + \Delta) 
    > \sigma - (\lambda - c) $, then 
    \begin{equation}
    \begin{aligned}
        \label{opt_deviation_large_sigma_1}
        F_\sigma(x) & = 
        \left\{\begin{aligned}
        & H(x), && \forall x \in [0, x^\dagger)\\ 
        & H(x^\dagger) , && \forall x \in [x^\dagger, \bar{x}_G)\\
        & H(\sigma+\Delta) , && \forall x \in [\bar{x}_G, x^\ddag)\\
        & 1 , && \forall x \in [x^\ddag, 1]\\
        \end{aligned}\right.
    \end{aligned}
    \end{equation}
    where $x^\dagger \in [0, x^*)$ satisfies 
    $\int_0^\sigma F_\sigma (x)dx = \sigma - (\lambda - c)$.
    
    \item if $\int_0^{x^*} H(x)dx + (\bar{x}_G - x^*) \cdot H(x^*) +
    (\sigma - \bar{x}_G) \cdot H(\sigma + \Delta) 
    \le \sigma - (\lambda - c) $, then 
        \begin{equation}
    \begin{aligned}
        \label{opt_deviation_large_sigma_2}
        F_\sigma(x) & = 
        \left\{\begin{aligned}
        & H(x), && \forall x \in [0, x^*)\\ 
        & H(x^*) , && \forall x \in [x^*, x')\\
        & H(\sigma+\Delta) , && \forall x \in [x', x^\ddag)\\
        & 1 , && \forall x \in [x^\ddag, 1]\\
        \end{aligned}\right.
    \end{aligned}
    \end{equation}
    where $x'\in[x^*, \bar{x}_G]$ satisfies 
    $\int_0^\sigma F_\sigma (x)dx = \sigma - (\lambda - c)$.
\end{enumerate}
\end{restatable}

\begin{proof}[Proof of \Cref{opt_deviation_structure_large_sigma}]
We first show the unique
existence of $\Delta$ such that 
$\sigma - (\lambda - c) + H(\sigma+\Delta) \cdot \Delta
= \int_0^{\sigma+\Delta} H(x)dx$. 
Fix $\sigma \in [\bar{x}_G, \sigma_H)$,
consider a function 
$f(x) := \sigma - (\lambda - c) + 
H(\sigma + x)\cdot x - \int_0^{\sigma+x} H(t)dt$.
Clearly, $f(\cdot)$ is 
continuously differentiable and increasing 
over $[0, 1 - \sigma]$. 
Note that 
\begin{align*}
    f(\sigma_H - \sigma) 
    & = (\sigma_H - \sigma) \cdot(H(\sigma_H) - 1) 
    \le 0\\
    f(1 - \sigma) 
    & =  \sigma - (\lambda - c) + 
    H(1)\cdot(1 - \sigma) - \int_0^{1}H(t)dt
    = c > 0~.
\end{align*}
Thus, there must exist 
a unique $\Delta \in (\sigma_H -\sigma, 1- \sigma)$
such that $f(\Delta) = 0$.
In below, we show the optimality 
of solution~\eqref{opt_deviation_large_sigma_1}
and~\eqref{opt_deviation_large_sigma_2}
via constructing a dual solution that satisfies
the complementary slackness conditions in
Equations \eqref{opt_CP_condi_1} and \eqref{opt_CP_condi_2}.
Fix a $\sigma \in [\bar{x}_G, \sigma_H)$,
and its corresponding $\Delta$. 
For notation simiplicity, we define
$p^\dagger := H(x^\dagger)^{n-1}$ in first case
and $p^* := H(x^*)^{n-1}$ in second case,
and $p_H := H(\sigma_H)^{n-1}$.
\squishlist
\item 
When 
$\int_0^{x^*} H(x)dx + (\bar{x}_G - x^*) \cdot H(x^*) +
(\sigma - \bar{x}_G) \cdot H(\sigma + \Delta) 
> \sigma - (\lambda - c)$,
in this case, 
let $\alpha_G := \frac{p_H - p^\dagger}{\bar{x}_G - x^\dagger}$,
and consider following dual solution
\begin{equation*}
\begin{aligned}
    \alpha 
    & = 
    -\frac{\alpha_G 
    \cdot (\sigma+\Delta - x^\dagger) + p^\dagger - p_H }{\Delta}~; \\
    p(x) 
    & = 
    \left\{\begin{aligned}
    & G(x)^{n-1} - \alpha  \cdot (\sigma - x), 
    && \forall x \in [0, x^\dagger)\\ 
    & \alpha_p \cdot (x - x^\dagger) + p^\dagger 
    - \alpha\cdot (\sigma - x^\dagger), 
    && \forall x \in [x^\dagger, \sigma+\Delta)\\ 
    & p_H, 
    && \forall x \in [\sigma+\Delta, 1]
    \end{aligned}\right.
\end{aligned}
\end{equation*}
where $\alpha_p := \alpha + \alpha_G$. 
We now show that the above 
constructed $p(\cdot)$ is global 
convex over $[0, 1]$, 
and $p(\cdot), \alpha$ satisfy the 
complementary slackness conditions 
in Equations~\eqref{opt_CP_condi_1} and \eqref{opt_CP_condi_2}.

To see the convexity of $p$, note that 
for any $x\in [0, x^\dagger]$, 
$\frac{\partial p(x)}{\partial x} = (G(x)^{n-1}) ' + \alpha$
is increasing due to the convexity 
$G^{n-1}$ over $[0, x^\dagger]$. 
Moreover, 
\begin{align*}
    \lim_{x\rightarrow (x^\dagger)^-} \frac{\partial p(x)}{\partial x}
    & = (G(x^\dagger)^{n-1})' + \alpha
    \le \alpha_p 
    = \alpha + \alpha_G~; \\
    \lim_{x\rightarrow (\sigma+\Delta)^-} \frac{\partial p(x)}{\partial x}
    & = -\frac{\alpha_G \cdot(\sigma - x^\dagger) - (p_H - p^\dagger)}{\Delta}
    = -(p_H - p^\dagger) \cdot \frac{\frac{\sigma - x^\dagger}{\bar{x}_G - x^\dagger} - 1}{\Delta} \le 0~.
\end{align*}
To check the continuity of $p$, 
note that 
\begin{align*}
    \lim_{x\rightarrow (x^\dagger)^-} p(x) 
    & =  G(x^\dagger)^{n-1} - \alpha \cdot (\alpha - x^\dagger)
    = p^\dagger - \alpha \cdot (\alpha - x^\dagger) = p(x^\dagger)~; \\
    \lim_{x\rightarrow (\sigma+\Delta)^-} p(x) 
    & = \alpha_p \cdot (\sigma+\Delta - x^\dagger) + p^\dagger - \alpha \cdot (\sigma - x^\dagger)\\
    & = \alpha \cdot \Delta + \alpha_G \cdot (\sigma+\Delta - x^\dagger)
    + p^\dagger
    = p_H ~.
\end{align*}
Thus, $p(\cdot)$ is convex over $[0, 1]$.

To satisfy the condition~\eqref{opt_CP_condi_2},
note that for $x\in[x^\dagger, \bar{x}_G)$, we have 
\begin{align*}
    p(x) + \alpha  \cdot (\sigma - x) - G(x)^{n-1}
    = ~ & (x - x^\dagger) (\alpha_p - \alpha) 
    - (G(x)^{n-1} - p^\dagger) \\
     = ~ & \alpha_G \cdot (x - x^\dagger)  - (G(x)^{n-1} - p^\dagger) 
    \labelrel{\ge}{convex_1} 0~,\\
    \Rightarrow ~~  
    p(x) + \alpha  \cdot (\sigma - x)  \ge~ & G(x)^{n-1}~,
\end{align*}
where \eqref{convex_1} is from the 
convexity of $G^{n-1}$
over $[0, \bar{x}_G)$.
Note $F_\sigma$ has non-zero 
support on $\bar{x}_G$.
For $x\in[\bar{x}_G, \sigma)$, we know
\begin{align*}
    p(\bar{x}_G) 
    & = p_H - \alpha (\sigma - \bar{x}_G)~;\\
    p(x) + \alpha  \cdot (\sigma - x) - G(x)^{n-1}
    & = \alpha_G \cdot (x - x^\dagger)  - (p_H  - p^\dagger) \ge 0 \\
    \Rightarrow ~~  
    p(x) + \alpha  \cdot (\sigma - x)  & \ge p_H~.
\end{align*}
For $x\in[\sigma, \sigma+\Delta]$, we 
already know $\alpha_p \le 0$ 
and $p(\sigma+\Delta) = p_H$, thus 
we have $p(x) \ge p_H, \forall x\in[\sigma, \sigma+\Delta]$.

Lastly, to satisfy condition~\eqref{opt_CP_condi_1},
as $F_\sigma(x) = H(x), \forall x\in[0, x^\dagger]$, 
it suffices to ensure 
\begin{align*}
    \int_{x^\dagger}^1 p(x)dF_\sigma(x) = \int_{x^\dagger}^1 p(x)dH(x)~.
\end{align*}
Now note that 
\begin{align}
    \int_{x^\dagger}^1 p(x)dF_\sigma(x)
    & = (H(\sigma+\Delta) - H(x^\dagger))\cdot p(\bar{x}_G)
    + (1-  H(\sigma+\Delta)) \cdot p_H ~.
    \label{condi_2_check_1}\\
    \int_{x^\dagger}^1 p(x)dH(x)
    & =  p_H - p(x^\dagger) \cdot H(x^\dagger) - \alpha_p \cdot 
    \int_{x^\dagger}^{\sigma+\Delta} H(x)dx~.
    \label{condi_2_check_2}
\end{align}
Consider
\begin{align*}
    & \eqref{condi_2_check_1} - \eqref{condi_2_check_2} \\
    \labelrel{=}{condi_2_helper_1} ~ &  -\alpha_p\cdot\left(H(x^\dagger) \cdot(\bar{x}_G - x^\dagger) + H(\sigma+\Delta)\cdot(\sigma+\Delta - \bar{x}_G) -  
    \int_{x^\dagger}^{\sigma+\Delta} H(x)dx\right)\\
    \labelrel{=}{condi_2_helper_2} ~ &   -\alpha_p\cdot \left(\sigma - (\lambda - c) + H(\sigma+\Delta) \Delta  - \int_0^{\sigma+\Delta} H(x)dx\right)
    \labelrel{=}{condi_2_helper_3} 0~,
\end{align*}
where \eqref{condi_2_helper_1} uses 
the definition of $p(\cdot)$
over $[x^\dagger, \sigma+\Delta]$,
\eqref{condi_2_helper_2} uses the definition
of $x^\dagger$, namely,
$\int_0^{x^\dagger} H(x)dx + (\bar{x}_G - x^\dagger) H(x^\dagger)
+ (\sigma - \bar{x}_G) H(\sigma+\Delta) = \sigma - (\lambda - c)$,
and \eqref{condi_2_helper_3} 
is from the definition of $\Delta$.

Putting all pieces together, we know 
the above $\alpha$, and $p$ is a dual solution
that satisfies the complementary slackness, 
leading the optimality of $F_\sigma$
in~\eqref{opt_deviation_large_sigma_1}.

\item 
When 
$\int_0^{x^*} H(x)dx + (\bar{x}_G - x^*) \cdot H(x^*) +
(\sigma - \bar{x}_G) \cdot H(\sigma + \Delta) 
\le \sigma - (\lambda - c)$,
in this case, 
let $\alpha_G := \frac{p_H - p^*}{\bar{x}_G - x^*}$,
i.e., the slope of the last 
linear portion of $G$, 
and consider following dual solution
\begin{equation}
\begin{aligned}
    \alpha & = 
    -\frac{\alpha_G 
    \cdot (\sigma+\Delta - x^*) + p^* - p_H }{\Delta}~; \\
    ~~
    p(x) & = 
    \left\{\begin{aligned}
    & G(x)^{n-1} - \alpha  \cdot (\sigma - x), 
    && \forall x \in [0, x^*)\\ 
    & \alpha_p \cdot (x - x^*) + p^* 
    - \alpha\cdot (\sigma - x^*), 
    && \forall x \in [x^*, \sigma+\Delta)\\ 
    & p_H, 
    && \forall x \in [\sigma+\Delta, 1]
    \end{aligned}\right.
\end{aligned}
\end{equation}
where $\alpha_p := \alpha_G + \alpha$.
Follow the analysis in earlier case, 
one can show that the above constructed 
$p$ is convex over $[0, 1]$, 
and $\alpha, p$ satisfy the complementary 
slackness conditions in~\eqref{opt_CP_condi_1}
and \eqref{opt_CP_condi_2}, showing 
that the solution in \eqref{opt_deviation_large_sigma_2}
is an optimal solution.
\squishend
The proof then completes.
\end{proof}

\begin{proof}[Proof of \Cref{opt_deviation_structure}]
We first prove the optimal 
structure of $F_\sigma$ 
for $\sigma\in[\sigma_{\NI}, \bar{x}_G]$.
We begin with analyzing following general problem
for any $\sigma \in [\sigma_{\NI}, \sigma_H)$,
\begin{equation}
\begin{aligned}
    \label{general_w_sigma}
    \max_{F \in \cH} ~ \int_0^1 u(x) dF(x) \quad
    \text{s.t.} ~ 
    \int_0^{\sigma} F(x) dx = \sigma - (\lambda - c)~.
\end{aligned}
\end{equation}
The above program has two major constraints, 
one is $F \in \cH$ 
to account for the feasibility of \infoStrategy\ $F$, 
and the other one accounts for
$\sigma_F = \sigma$ (recall \Cref{lem_sigma_defn}).
The above optimization problem is non-trivial 
as sender $i$ can deviate to any possible 
strategy $F\in\cH(\sigma)$,
and this is an infinite-dimensional linear program. 
Nevertheless, some recent technical developments 
in the information design literature are 
useful to our problem. 
In particular, we use the following result 
obtained by \citet{DM-19}, which provides
a duality theory for optimization problems 
with MPS constraints.
To be more precise, they
consider the problem 
$\max_{F: H \blackwellOrder F} \int_0^1 u(x) dF(x)$, and 
show that if $F$ is the solution to this program,
then there must exist a convex function $p(x): 
[0, 1] \rightarrow \R$ such 
\begin{align}
    \label{opt_CP_condi_1}
    \int_0^1 p(x)d F(x) = \int_0^1 p(x)d H(x)~,
\end{align}
and $F$ is also the optimal solution 
to the program 
$\max_{\widetilde{F} \in \Delta([0, 1])} 
\int_0^1(u(x) - p(x)) d\widetilde{F}(x)$.
In our problem, additional to 
the MPS constraint, 
we also have a linear constraint that the 
strategy $F$ has $\sigma_F = \sigma$.
Follow the similar analysis, one can deduce 
that if $F_\sigma$ is the optimal solution
to the program~\eqref{general_w_sigma}, it must 
also exist a convex function $p(\cdot)$
where~\eqref{opt_CP_condi_1} holds for $F_\sigma$, 
and there exists $\alpha \in \R$ such that
\begin{align*}
    F_\sigma \in \argmax_{\widetilde{F} \in \Delta([0, 1])} & \left\{
    \int_0^1 (u(x) - p(x)) d\widetilde{F}(x) -  \alpha \cdot \left(\sigma \int_0^{\sigma} d\widetilde{F}(x) - \int_0^{\sigma} xd\widetilde{F}(x) - \sigma + (\lambda - c)\right)\right\}~,
\end{align*}
where we have used integration by parts 
in the reservation value constraint. 
Observe that we can always add a constant to 
$p(\cdot)$ without changing any of its properties.
Thus, by complementary slackness, one must have
\begin{equation}
\begin{aligned}
    \label{opt_CP_condi_2}
    \text{if } x\in [0, \sigma) \wedge x\in\supp{F_\sigma}: 
    ~~ & u(x) = p(x) + \alpha\cdot(\sigma - x) \\
    \text{if } x\in [0, \sigma) \wedge x\notin\supp{F_\sigma}: 
    ~~ & u(x) \le p(x) + \alpha\cdot(\sigma - x) \\
    \text{if } x\in [\sigma, 1] \wedge x\in\supp{F_\sigma}: 
    ~~ & u(x) = p(x) \\
    \text{if } x\in [\sigma, 1]  \wedge x\notin\supp{F_\sigma}: 
    ~~ & u(x) \le p(x)~.
\end{aligned}
\end{equation}
Now to prove the optimal solution 
defined as in~\eqref{opt_deviation}, 
it suffices to show that there exists 
a convex function $p(\cdot)$
and a value $\alpha\in\R$ that 
satisfies the conditions 
in~\eqref{opt_CP_condi_1} and~\eqref{opt_CP_condi_2}
with $u(x) = \min\left\{G(x)^{n-1}, 
G(\sigma)^{n-1}\right\}$.
We consider
\begin{equation*}
\begin{aligned}
    \alpha = 
    -\frac{G(\sigma)^{n-1} - G(x^\dagger)^{n-1}}{\sigma - x^\dagger}~; 
    \quad\quad
    p(x) = 
    \left\{\begin{aligned}
    & G(x)^{n-1} - \alpha  \cdot (\sigma - x), 
    && \forall x \in [0, x^\dagger)\\ 
    & G(\sigma)^{n-1}, 
    && \forall x \in [x^\dagger, 1]
    \end{aligned}\right.
\end{aligned}
\end{equation*}
To check the convexity of $p$, 
note that 
$\frac{\partial p(x)}{\partial x} 
= \frac{\partial G(x)^{n-1}}{\partial x} + \alpha$ 
is increasing 
over $[0, x^\dagger]$ since $G^{n-1}$ 
is convex over $[0, x^\dagger]$.
Moreover, 
$\frac{\partial p(x^\dagger)}{\partial x^\dagger} 
= (G(x^\dagger)^{n-1})' + \alpha \le 0$
as $G^{n-1}$ is convex over $[0, \sigma]$, 
and $\lim_{x\rightarrow (x^\dagger)^-} p(x) 
= G(\sigma)^{n-1}$. 
Thus, $p(\cdot)$ is global convex over $[0, 1]$.

To satisfy the condition~\eqref{opt_CP_condi_2},
note for $x\in[x^\dagger, \sigma]$, we have 
\begin{align*}
    & p(x) + \alpha  \cdot (\sigma - x) - G(x)^{n-1} \\
    & =  (\sigma - x) \cdot \left(\frac{G(\sigma)^{n-1} - G(x)^{n-1} }{\sigma - x } -\frac{G(\sigma)^{n-1} - G(x^\dagger)^{n-1}}{\sigma - x^\dagger} \right)~.
\end{align*}
Thus, we have $p(x) + \alpha  \cdot (\sigma - x) \ge G(x)^{n-1}$.
Together with $p(x) = G(\sigma)^{n-1}, \forall x\in[\sigma, 1]$, 
we know that $p(\cdot)$ satisfies the condition~\eqref{opt_CP_condi_2}. 

Lastly, to satisfy the condition~\eqref{opt_CP_condi_1},
as $F_\sigma(x) = H(x), \forall x\in[0, x^\dagger]$, 
it suffices to ensure 
\begin{align*}
    \int_{x^\dagger}^1 p(x)dF_\sigma(x) = \int_{x^\dagger}^1 p(x)dH(x)~,
\end{align*}
where the above holds true as they both equal to 
$G(\sigma)^{n-1}\cdot(1 - H(x^\dagger))$.
Thus the constructed $p$ and $\alpha$ 
satisfy the conditions 
in~\eqref{opt_CP_condi_1}--\eqref{opt_CP_condi_2},
implying the solution in~\eqref{opt_deviation} 
is an optimal solution.

With the above characterized $F_\sigma$, 
we now prove the second part of the above result, 
i.e., $\OPT_\sigma$ is monotone increasing
w.r.t. $\sigma\in[\sigma_{\NI}, \bar{x}_G]$.
By definition, we have
\begin{align}
    \OPT_\sigma
    = \int_0^{x^\dagger} G(x)^{n-1}dH(x) + 
    G(\sigma)^{n-1}\cdot(1 - H(x^\dagger))~.
    \label{eq_opt_deviation_case_2}
\end{align}
Recall that $x^\dagger$ satisfies
$\int_0^{x^\dagger} H(x) dx + (\sigma - x^\dagger) 
\cdot H(x^\dagger) = \sigma - (\lambda - c)$,
thus, 
$\sigma = \frac{\int_0^{x^\dagger} H(x) dx- x^\dagger H(x^\dagger) 
+ (\lambda -c)}{1 - H(x^\dagger)}.$
Define a function 
$\sigma(x) 
:= \frac{\int_0^{x} H(t) dt- x H(x)+ (\lambda -c)}{1 - H(x)}$.
Now back to~\eqref{eq_opt_deviation_case_2},
we have
\begin{align*}
    \OPT_\sigma
    = \int_0^{x^\dagger} G(x)^{n-1}dH(x) + 
    G(\sigma(x^\dagger))^{n-1}\cdot(1 - H(x^\dagger))~.
\end{align*}
Consider a function $f(x) := \int_0^{x} G(t)^{n-1}dH(t) + 
    G(\sigma(x))^{n-1}\cdot(1 - H(x))$.
Let $g(\cdot)$ denote the density 
function of distribution $G$.
Now observe that 
\begin{align*}
    \frac{\partial f(x)}{\partial x}
    & = G(x)^{n-1}  h(x) + (n-1)  G(\sigma(x))^{n-2}  g(\sigma(x)) \sigma(x)'  (1 - H(x)) - G\left(\sigma(x)\right)^{n-1}  h(x)\\
    & = h(x) \cdot \bigg(\left(G(x)^{n-1} - G\left(\sigma(x)\right)^{n-1}\right)
    + \frac{\partial G(\sigma(x))^{n-1}}{\partial \sigma(x)}
    \cdot \left(\sigma(x)-x\right)\bigg)
    \labelrel{\ge}{helper_convexity} 0~,
\end{align*}
where in~\eqref{helper_convexity}, we use the convexity 
of $G^{n-1}$ over its support in $[0, \bar{x}_G]$, 
and $\sigma(x) \ge x, \forall x\in[0, \bar{x}_G]$,
and $h(x) \ge 0, \forall x$.
This implies that the optimal deviation
payoff is increasing w.r.t. $x^\dagger$, 
and thus increasing w.r.t. 
$\sigma \in [\sigma_{\NI}, \bar{x}_G]$.
\end{proof}

To prove \Cref{opt_deviation_payoff},
we first show the following monotonicity result.
\begin{claim}
\label{monotonicity_Delta}
Fix a $\sigma\in(\bar{x}_G, \sigma_H)$
and its corresponding $\Delta$ such that 
$\sigma - (\lambda - c) + H(\sigma+\Delta) \cdot \Delta
= \int_0^{\sigma+\Delta} H(x)dx$.
When $\sigma$ increases, 
the value $\sigma+\Delta$ will decrease. 
\end{claim}
\begin{proof}[Proof of \Cref{monotonicity_Delta}]
To prove the above result,
consider a function $\nu(\sigma, y) 
:= \sigma - (\lambda - c) + H(y)\cdot(y - \sigma) - 
\int_0^y H(t)dt$.
Clearly 
$\frac{\partial \nu(\sigma, y)}{\partial \sigma}
= 1 - H(y) \ge 0$
and $\frac{\partial \nu(\sigma, y)}{\partial y}
= H(y) + h(y)(y-\sigma) - H(y) \ge 0$ for $y \ge \sigma$.
Consider $\sigma_1, \sigma_2$
where $\sigma_1 < \sigma_2$, and their
corresponding $\Delta_1, \Delta_2$ such that
$\nu(\sigma_1, \sigma_1 + \Delta_1) = 0$
and $\nu(\sigma_2, \sigma_2 + \Delta_2) = 0$.
Then by monotonicity of $\tau(\sigma, \cdot)$
and $\tau(\cdot, y)$, we have
\begin{align*}
    \tau(\sigma_2, \sigma_2 + \Delta_2) = 0 
    = \tau(\sigma_1, \sigma_1 + \Delta_1) \le
    \tau(\sigma_2, \sigma_1 + \Delta_1)
    ~ \Rightarrow ~
    \sigma_2 + \Delta_2 \le \sigma_1 + \Delta_1~. 
\end{align*}
\end{proof}

We are now ready to present our proof
for \Cref{opt_deviation_payoff}.
\begin{proof}[Proof of \Cref{opt_deviation_payoff}]
We consider following possible cases
based on the value of 
$\sigma_{\NI} = \lambda - c$ and $\bar{x}_G$.
\begin{itemize}
    \item
    When $\lambda - c \ge \bar{x}_G$, 
    we know that $\sigma > \bar{x}_G, 
    \forall \sigma\in[\sigma_{\NI}, \sigma_H)$.
    Thus, for any $\sigma \in [\sigma_{\NI}, \sigma_H)$,
    the optimal deviation $F_\sigma$ follows the 
    characterizations in \Cref{opt_deviation_structure_large_sigma}.
    Fix a $\sigma$ and its corresponding 
    $\Delta$ where 
    $\sigma - (\lambda - c) + 
    H(\sigma + \Delta)\cdot \Delta 
    = \int_0^{\sigma+\Delta} H(t)dt$.

    \textbf{In first case of \Cref{opt_deviation_structure_large_sigma}},
    with structure of $F_\sigma$, we can 
    write the payoff of deviating to $F_\sigma$
    as follows:
    \begin{align}
        \label{eq_opt_large_sigma_1}
        \OPT_\sigma
        = \int_0^{x^\dagger} G(x)^{n-1}dH(x) + 
        H(\sigma_H)^{n-1}\cdot(1 - H(x^\dagger))~.
    \end{align}
    We will now show that $\OPT_\sigma$ 
    is decreasing w.r.t 
    $\sigma \in [\sigma_{\NI}, \sigma_H)$.
    Recall that $x' = x^\dagger$ satisfies
    \begin{align*}
        \int_0^{x^\dagger} H(x)dx + (\bar{x}_G - x^\dagger)
        H(x^\dagger) + (\sigma - \bar{x}_G) H(\sigma+\Delta)
        = \sigma - (\lambda - c)~.
    \end{align*}
    Thus, with the definition of $\Delta$, we have
    \begin{align*}
        H(\sigma+\Delta)\cdot  (\sigma+\Delta - \bar{x}_G) 
        + H(x^\dagger) \cdot(\bar{x}_G - x^\dagger)
        = \int_{x^\dagger}^{\sigma+\Delta} H(t)dt~.
    \end{align*}
    Now consider following function 
    $\tau: [\sigma, 1] \times [0, \bar{x}_G] \rightarrow \R$
    \begin{align*}
        \tau(y, x)
        := H(y) \cdot (y - \bar{x}_G) + 
        H(x) \cdot(\bar{x}_G - x) 
        - \int_x^y H(t)dt~.
    \end{align*}
    Clearly, we have
    \begin{align*}
        \frac{\partial \tau(y, x)}{\partial y}
        = h(y)\cdot(y-\bar{x}_G) \ge 0; ~~
        \frac{\partial \tau(y, x)}{\partial x}
        =  h(x)(\bar{x}_G -x) \ge 0 ~.
    \end{align*}
    Consider $\sigma_1, \sigma_2$
    where $\sigma_1 < \sigma_2$, and their corresponding
    $\Delta_1, \Delta_2$, $x^\dagger_1, x^\dagger_2$, 
    such that $\tau(\sigma_1 + \Delta_1, x^\dagger_1) = 0$
    and $\tau(\sigma_2 + \Delta_2, x^\dagger_2) = 0$
     Then by monotonicity of $\tau(y, \cdot)$
    and $\tau(\cdot, x)$, we have
    \begin{align*}
        \tau(\sigma_2 + \Delta_2, x^\dagger_2) = 0 
        = \tau(\sigma_1 + \Delta_1, x^\dagger_1)
        \ge \tau(\sigma_2 + \Delta_2, x^\dagger_1)
        ~ \Rightarrow ~
        x^\dagger_2 \ge x^\dagger_1~,
    \end{align*}
    where we have used the 
    result in \Cref{monotonicity_Delta}.
    Thus, we have showed that
    when $\sigma$ increases, the value 
    $x^\dagger$ will also increase. 
    
    Now back to \eqref{eq_opt_large_sigma_1},
    consider a function
    $f(x) := \int_0^{x} G(t)^{n-1}dH(t) + 
    H(\sigma_H)^{n-1}\cdot(1 - H(x))$,
    then $\forall x\in[0, \bar{x}_G]$, 
    \begin{align*}
        \frac{\partial f(x)}{\partial x}
        = G(x)^{n-1}h(x) - H(\sigma_H)^{n-1}h(x)
        = h(x)\cdot(G(x)^{n-1} - H(\sigma_H)^{n-1}) \le 0~,
    \end{align*}
    implying that $f(x)$
    is strictly decreasing w.r.t $x\in[0, \bar{x}_G]$.
    Consequently, we have showed
    that the value $\OPT_\sigma$ 
    is decreasing w.r.t $\sigma$.

    
    \textbf{In second case of \Cref{opt_deviation_structure_large_sigma}},
    we have
    \begin{align}
        \label{eq_opt_large_sigma_2}
        \OPT_\sigma = \int_0^{x^*} H(x)dx + (H(\sigma+\Delta) - H(x^*)) \cdot G(x_1)^{n-1} + H(\sigma_H)^{n-1}\cdot(1 - H(\sigma+\Delta))~,
    \end{align}
    where $x_1$ satisfies that
    $x_1 
    = \frac{\int_0^{x^*} H(x)dx - x^* H(x^*) + (\sigma+\Delta) H(\sigma+\Delta) - \int_0^{\sigma+\Delta} H(t)dt}{H(\sigma+\Delta) - H(x^*)}$.
    Recall that $\Delta \in (\sigma_H - \sigma, 1-\sigma)$,
    and $x_1\in[x^*, \bar{x}_G]$.
    Define a function 
    $\kappa(x): [\sigma_H, 1] \rightarrow [x^*, \bar{x}_G]$
    \begin{align*}
        \kappa(x) := \frac{\int_0^{x^*} H(x)dx - x^* H(x^*) + xH(x) - \int_0^x H(t)dt}{H(x) - H(x^*)}~.
    \end{align*}
    Now back to~\eqref{eq_opt_large_sigma_2}
    and consider following function
    $f: [\sigma_H, 1] \rightarrow \R$:
    \begin{align*}
        f(x) := \int_0^{x^*} H(t)dt + (H(x) - H(x^*)) \cdot G(\kappa(x))^{n-1} + H(\sigma_H)^{n-1} (1 - H(x))~. 
    \end{align*}
    Observe that
    \begin{align}
        \frac{\partial f(x)}{\partial x} 
        =  h(x) \cdot\left(G(\kappa(x))^{n-1} - H(\sigma_H)^{n-1}) + \frac{\partial G(\kappa(x))^{n-1}}{\partial \kappa(x)} \cdot (x - \kappa(x))\right) \ge 0~.
        \label{opt_monotonicity_2}
    \end{align}
    Recall that in \Cref{monotonicity_Delta}, 
    we have showed larger $\sigma$ 
    will induce smaller $\sigma+\Delta$. 
    Together with~\eqref{opt_monotonicity_2}, 
    we can conclude that the value $\OPT_\sigma$
    is decreasing w.r.t $\sigma$.
    
    Combined with the earlier analysis
    for the first case of \Cref{opt_deviation_structure_large_sigma},
    we can conclude that 
    \begin{align*}
        \max_{\sigma: \sigma\in[\sigma_{\NI}, \sigma_H)}
        \OPT_\sigma = \OPT_{\sigma_\NI}
        = G(\sigma_\NI)^{n-1} = G(\lambda - c)^{n-1}~.
    \end{align*}
    Thus, to ensure $\OPT_\sigma \le\sfrac{1}{n}$,
    it suffices to ensure 
    $G(\lambda - c)^{n-1} \le \sfrac{1}{n}$.

    \item 
    When $\lambda - c < \bar{x}_G$.
    Follow the analysis in case $(i)$,
    for any $\sigma \in [\bar{x}_G, \sigma_H)$,
    we know 
    \begin{align*}
        \OPT_\sigma \le \OPT_{\bar{x}_G}~.
    \end{align*}
    Now consider the deviation $F$ which 
    satisfies $\sigma_F \in [\sigma_{\NI}, \bar{x}_G]$,
    from the proof for \Cref{opt_deviation_structure}, 
    we know 
    \begin{align}
        \max_{\sigma\in[\sigma_{\NI}, \bar{x}_G]} \OPT_\sigma
        = \OPT_{\bar{x}_G} 
        = \int_0^{x^\dagger} G(x)^{n-1}dH(x) + 
        H(\sigma_H)^{n-1}(1 - H(x^\dagger)) ~,
        \label{eq_deviation_payoff_general_Case2}
    \end{align}
    where $x^\dagger$ satisfies 
    $\int_0^{x^\dagger} H(x) dx + 
    (\bar{x}_G - x^\dagger) \cdot H(x^\dagger) 
    = \bar{x}_G - (\lambda - c)$, i.e., 
    $\int_{x^\dagger}^1 (x - \bar{x}_G)dH(x)=c$.
    As a result, 
    to ensure $\OPT_\sigma\le \sfrac{1}{n}$, 
    it suffices to ensure 
    $\eqref{eq_deviation_payoff_general_Case2} 
    \le \sfrac{1}{n}$.
\end{itemize}
\end{proof}

Combine the above results, we now prove our main theorem.
\begin{proof}[Proof of \Cref{thm_nash_iff}]
For the ``if'' direction, it suffices to 
show that no sender has profitable deviation 
under the strategy profile $(G, \ldots, G)$
where $G$ satisfies
conditions $(i)$--$(iii)$ in \Cref{thm_nash_iff}. 
Consider following
two kinds of deviations: one is deviating 
to a strategy $F$ where $\sigma_F = \sigma_H$, 
i.e., $F\in\cH(\sigma_H)$,
and the other is deviating 
to a strategy $F$ where $\sigma_F = \sigma < \sigma_H$, 
i.e., $F\in\cH(\sigma)$.
From the first part of \Cref{symme_deviation}, 
we know there is no such
profitable deviation to a strategy $F\in\cH(\sigma_H)$.
From \Cref{lem_most_profitable_devia} 
and \Cref{opt_deviation_structure},
we know there is no such profitable deviation 
to a strategy $F\in\cH(\sigma), 
\forall \sigma < \sigma_H$.
Thus, $(G, \ldots, G)$
must be an equilibrium.
For the ``only if'' direction, 
\Cref{max_sigma_nash} proves the condition $(i)$.
The condition $(ii)$ follows from the second
part of \Cref{symme_deviation}. 
The conditions $(iii)$ follows from 
the definition of equilibrium.
Namely, it is not profitable to deviate to a 
strategy that has the reservation value 
$\max\{\sigma_\NI, \bar{x}_G\}$, thus the optimal 
deviation value is no larger than $\sfrac{1}{n}$, 
with \Cref{opt_deviation_structure}, this
is exactly the statement of the condition $(iii)$.
\end{proof}

\section{Missing proofs of \Cref{sec-implications}}
\label{proof-sec-implications}

\begin{proof}[Proof of \Cref{full_infor_nash_iif}]
When $H^{n-1}$ is convex over $[0, \sigma_H]$,
it is easy to see that the unique distribution 
$G$ that meets conditions $(i)$--$(ii)$
in \Cref{thm_nash_iff} must satisfy that
$G(x) = H(x), \forall x\in[0, \sigma_H]$.
We now show how the condition $(iii)$ always 
holds when $H^{n-1}$ is convex over $[0, \sigma_H]$. 
In this case, we know $\bar{x}_G = \sigma_H$,
and $\sigma_{\NI} = \lambda - c 
< \sigma_H = \bar{x}_G$, thus, it suffices to 
show the case $(b)$ in condition $(iii)$ holds.
Clearly, when $\bar{x}_G = \sigma_H$, we have $x^\dagger = \bar{x}_G$, and
\begin{align*}
    & \int_0^{x^\dagger} G(x)^{n-1}dH(x) + 
    H(\sigma_H)^{n-1}(1 - H(x^\dagger)) \\
    & = \frac{1}{n} \cdot H(\sigma_H)^n + 
    H(\sigma_H)^{n-1}(1 - H(\sigma_H)) 
    \le \frac{1}{n}~,
\end{align*}
where the last inequality always holds by algebra
for any $n \ge 2$. 
Thus, $G$, i.e., the essentially full information
disclosure, is the equilibrium strategy.
\end{proof}

\begin{proof}[Proof of \Cref{large_n_convex}]
Consider the second-order derivative of function $H^{n-1}$:
\begin{align*}
    \frac{\partial^2 H^{n-1}(x)}{\partial x^2}
    = (n-1) H^{n-3}(x) \left((n-2)h(x)^2 + H(x)h'(x)\right)
\end{align*}
where $h(x), h'(x)$ are the first-order, second-order derivative 
of the prior $H$, respectively. 
As we can see, when $n$ is large enough, one can ensure that the right-hand side
of the above equality is always larger than $0$, which 
guarantees the convexity of the function $H^{n-1}$.
\end{proof}

\begin{proof}[Proof of \Cref{essential_full_payoff}]
Recall that from \Cref{agent_opt_welfare},
we know under essentially full information equilibrium, 
we have
$\agentU(G, \ldots, G) 
= \sigma_H - \int_0^{\sigma_H} H(x)^n dx~.$
Consider function $f(x, n) := x - \int_0^{x} H(t)^n dt$.
Clearly, we have $\frac{\partial f(x, n)}{\partial x} 
= 1 - H(x)^n > 0$. 
Thus, agent's payoff under essentially 
full information equilibrium
is strictly increasing w.r.t. $\sigma_H$. 
This implies that agent's payoff is decreasing 
w.r.t. the cost.
On the other hand, 
when $n$ increases, we have $H^n$ is more convex and the 
integral $\int_0^{x} H(t)^n dt$ is smaller, implying that
agent's payoff is increasing.
\end{proof}

\begin{proof}[Proof of \Cref{concave_Nash_iif}]
When $H^{n-1}$ is concave over $[0, \sigma_H]$,
it is easy to see that the unique distribution 
$G$ that meets condition $(i)$--$(ii)$,
must be that $G^{n-1}$ is 
linear over $[0, \bar{x}_G]$,
and $G$ has no support over $[\bar{x}_G, \sigma_H]$.
If $\lambda - c \ge \bar{x}_G$, then 
$G$ is equilibrium strategy
if and only if $G(\lambda -c)^{n-1} \le\sfrac{1}{n}$.
If $\lambda - c < \bar{x}_G$, we now show that
the case $(b)$ in condition $(iii)$
is equivalent to ensure $G(\lambda -c)^{n-1} 
\le\sfrac{1}{n}$.
To see this, let 
$k:= \frac{H(\sigma_H)^{n-1}}{\bar{x}_G}$
denote the slope
of the linear portion of $G^{n-1}$.
Then, for $x^\dagger$ satisfying
$\int_{x^\dagger}^1 (x-\bar{x}_G)dH(x)=c$, 
i.e., $\int_0^{x^\dagger} H(x)dx + (\bar{x}_G - x^\dagger)H(x^\dagger) = \bar{x}_G - (\lambda - c)$, 
note that
\begin{align*}
    & \int_0^{x^\dagger} G(x)^{n-1}dH(x) + 
    H(\sigma_H)^{n-1}(1 - H(x^\dagger)) \\
    = ~ & G(x^\dagger)^{n-1} H(x^\dagger) - k\int_0^{x^\dagger}H(x)dx+ 
    H(\sigma_H)^{n-1}(1 - H(x^\dagger))\\
    = ~ & k \cdot (\lambda - c) = G(\lambda -c)^{n-1}~,
\end{align*}
where we have used the linearity of 
$G^{n-1}$ over $[0, \bar{x}_G]$.
Thus, combining above two cases,
to guarantee $G$ is the equilibrium strategy, 
it suffices to ensure 
$G(\lambda - c)^{n-1} \le\sfrac{1}{n}$.
\end{proof}

\section{Missing proofs of Section \ref{sec-prelim}}
\label{sec-proof-of-prelim}

\begin{proof}[Proof of \Cref{agent_opt_welfare}]
Recall that from \Cref{thm_r_utility_sigma}, we know
\begin{align}
    \agentU(H_1, \ldots, H_n) = \max_{G_i': H_i\blackwellOrder G_i', \forall i} 
    ~ \agentU(G_1', \ldots, G_n')~.
\end{align}
Let us fix all boxes' strategies $G_{-i}' = (G_j')_{j\in[n] \setminus\{i\}}$.
Note that from \Cref{agent_opt},
\begin{align*}
   \expect[G_1', \ldots, G_n']{\max_i \kappa_{G_i'}}
   = \expect[G_{-i}']{\expect[G_i']{\max \left\{\kappa_{G_i'}, \kappa_{G_{-i}'}\right\}}}~,
\end{align*}
where $\kappa_{G_{-i}'} := \{\kappa_{G_1'}, \ldots, \kappa_{G_{i-1}'}, 
\kappa_{G_{i+1}'}, \ldots, \kappa_{G_n'}\}$.
For every possible $\kappa_{G_{-i}'} = b$, we have
$\expect[x\sim G_i']{\max \left\{\kappa_{G_i'}, b\right\}}
= 
\expect[x\sim G_i']{\max \left\{\min\{x_i, \sigma_{G_i'}\}, b\right\}}$.
Notice that when $b \le \sigma_{G_i'}$, we have
\begin{align*}
    \expect[x\sim G_i']{\max \left\{\min\{x_i, \sigma_{G_i'}\}, b\right\}}
    & = \int_{\sigma_{G_i'}}^1 \sigma_{G_i'} d G_i'(x) 
    + \int_b^{\sigma_{G_i'}} b dG_i'(x) 
    + \int_0^b xdG_i'(x)\\
    & = \sigma_{G_i'}(1 - G_i'(\sigma_{G_i'})) + b G_i'(\sigma_{G_i'}) - \int_0^b G_i'(x)dx~.
\end{align*}
When $b > \sigma_{G_i'}$, we have 
\begin{align*}
    \expect[x\sim G_i']{\max \left\{\min\{x_i, \sigma_{G_i'}\}, b\right\}} = b~.
\end{align*}
Recall that under essentially 
full information strategy $G_i$ for box $i$, we have 
$\sigma_{G_i} = \sigma_{H_i}, G_i(\sigma_{G_i}) = H_i(\sigma_{H_i})$, and 
$G_i(x) = H_i(x), \forall x\in[0, \sigma_{H_i}]$. 
Thus, for any $b$, we have 
\begin{align}
    \expect[x\sim G_i]{\max \left\{\min\{x_i, \sigma_{G_i}\}, b\right\}}
    = \expect[x\sim H_i]{\max \left\{\min\{x_i, \sigma_{H_i}\}, b\right\}}~,
\end{align}
which gives us for any $G_{-i}'$, we have 
$\expect[G_i, G_{-i}']{\max\left\{\kappa_{G_i}, \kappa_{G_{-i}'} \right\}} = \expect[H_i, G_{-i}']{\max\left\{\kappa_{H_i}, \kappa_{G_{-i}'} \right\}}$, implying
$\agentU(G_i, G_{-i}') = \agentU(H_i, G_{-i}')$. Similarly 
arguments can be carried over to all boxes' strategies. 
Thus, for an essentially full information strategy profile $G_1, \ldots, G_n$, 
we have $\agentU(G_1, \ldots, G_n) = \agentU(H_1, \ldots, H_n)$.

When $H \equiv H_i, \forall i\in[n]$, from \Cref{agent_opt}, we know
\begin{align*}
    \agentU(H, \ldots, H) 
    & =  \expect[x_i\sim H, \forall i]{\max\big\{\min\{x_1, \sigma_H\}, \ldots, \min\{x_n, \sigma_H\}\big\}}\\
    & = \sigma_H \cdot (1 - H(\sigma_H)^n) + \int_0^{\sigma_H} x d H(x)^n 
    = \sigma_H - \int_0^{\sigma_H} H(x)^n dx~.
\end{align*}
For an essentially full information disclosure strategy $G$,
we have $G(x) = H(x), \forall x\in[0, \sigma_H]$. 
Thus, 
\begin{align}
    \label{eq_r_opt_payoff}
    \agentU(G, \ldots, G) 
    = \sigma_H - \int_0^{\sigma_H} G(x)^n dx
    \labelrel{=}{cor_3_10_helper} \sigma_H - \int_0^{\sigma_H} H(x)^n dx
    = \agentU(H, \ldots, H)~,
\end{align}
where~\eqref{cor_3_10_helper} is from the definition
of strategy $G$.
\end{proof}

\begin{proof}[Proof of \Cref{prop_informativness_sigma}]
When inspection cost $c = 0$, we have the reservation values
$\sigma_G = \sigma_{G'} = +\infty$.
Below we prove the result for cost $c > 0$.
From \Cref{lem_sigma_defn}, we know
\begin{align*}
    \sigma_{G'} - \sigma_G 
    \labelrel{=}{helper_sec_3_1} 
    \int_{\sigma_G}^1 G(x)dx - \int_{\sigma_{G'}}^1 G'(x)dx
    \labelrel{\ge}{helper_sec_3_2}  
    \int_{\sigma_G}^1 G(x)dx - \int_{\sigma_{G'}}^1 G(x)dx~,
\end{align*}
where equality \eqref{helper_sec_3_1} is due to the definition 
$G' \blackwellOrder G$ which implies that 
$\expect[x\sim G']{x} = \expect[x\sim G]{x}$,
ineqaulity \eqref{helper_sec_3_2} is due to \Cref{MPS_defn}.
Now suppose $\sigma_{G'} < \sigma_G$, 
\begin{align*}
    \sigma_G - \sigma_{G'} 
    \le \int_{\sigma_{G'}}^{\sigma_G} G(x)dx 
    \labelrel{\le}{helper_sec_3_3}
    \sigma_G - \sigma_{G'},
\end{align*}
where inequality \eqref{helper_sec_3_3} holds only when 
$G(x) = 1, \forall x\in[\sigma_{G'}, \sigma_G]$. 
However we note that it cannot be $G(\sigma_{G'}) = 1$ 
when $\sigma_{G'} < \sigma_G$. 
Suppose $G(\sigma_{G'}) = 1$ when $\sigma_{G'} < \sigma_G$, 
then we have $G(\sigma_G) = 1$ and 
$\E_{x\sim G}[(x - \sigma_G)_+] = 0 \neq c$.
As a result, when $G(\sigma_{G'}) < 1$, we have 
$\sigma_G - \sigma_{G'}
\le \int_{\sigma_{G'}}^{\sigma_G} G(x)dx 
< \sigma_G - \sigma_{G'}$,
which contradicts itself. 
Thus, we must have $\sigma_{G'} \ge \sigma_G$.
\end{proof}

\begin{proof}[Proof of \Cref{interval_MPS}]
The condition for $\sigma_{G} = \lambda_i - c_i$
is straightforward from \Cref{lem_sigma_defn}. 
We next prove the condition for $\sigma_{G} = \sigma_{H_i}$.
For the ``if'' direction, 
note that from the definition of $H_i\blackwellOrder_{[0, \sigma_{H_i}]} G$,
we know $W(\sigma_{H_i}) = 0$, i.e., 
$\int_0^{\sigma_{H_i}} H_i(x) = \int_0^{\sigma_{H_i}} G(x)$, thus 
$\int_0^{\sigma_{H_i}} G(x) = \sigma_{H_i} - (\lambda_i -c_i)$.
From \Cref{lem_sigma_defn}, 
we then know $\sigma_G = \sigma_{H_i}$.
For the ``only if'' direction, 
from $\sigma_G = \sigma_{H_i}$, we know 
$\int_0^{\sigma_{H_i}} G(x) = \sigma_{H_i} - (\lambda_i - c_i)$, thus 
$\int_0^{\sigma_{H_i}} G(x)dx =\int_0^{\sigma_{H_i}} H_i(x)dx $,
implying $W(\sigma_{H_i}) = 0$.
As $H_i\blackwellOrder G$, we know $W(y) \ge 0, 
\forall y\in[0, \sigma_{H_i}]$.
Thus, $H_i\blackwellOrder_{[0, \sigma_{H_i}]} G$.
\end{proof}

\begin{restatable}{lemma}{lemsigmadefn}
\label{lem_sigma_defn}
For any $G$ with mean $\lambda$ 
and for any $c\ge 0$, 
$\sigma_G = \sigma$ if and only if 
$\int_0^{\sigma} G(x)dx = \sigma - (\lambda - c)$.
\end{restatable}
\begin{proof}[Proof of \Cref{lem_sigma_defn}]
By definition, we have
\begin{align*}
 c = \expect[x\sim G]{\max\{x - \sigma_G,0\}}
 = \int_{\sigma_G}^1(x - \sigma_G)dG(x) 
 = \lambda   + \int_0^{\sigma_G} G(x)dx - \sigma_G~,
\end{align*}
where we have used the fact $\int xdG(x) = \lambda$ 
and integral by parts.
Rearranging the terms gives us the result.
\end{proof}

\begin{restatable}{lemma}{lemsamedensity}
\label{lem_same_density}
For any $H$, a \infoStrategy\ $G: H \blackwellOrder G$  
satisfying $\sigma_G = \sigma_H$ must have 
$G(\sigma_H) = H(\sigma_H)$.
\end{restatable}
\begin{proof}[Proof of \Cref{lem_same_density}]
Recall that if $G$ satisfies $\sigma_G = \sigma_H$, 
from \Cref{lem_sigma_defn}, we have
$\int_0^{\sigma_H} G(x)dx = \sigma_H - (\lambda - c)
= \int_0^{\sigma_H} H(x)dx$.
We now consider following two possible cases:
\begin{itemize}
    \item 
    Suppose that $G(\sigma_H) > H(\sigma_H)$, 
    as $H$ is continuous over $[0, 1]$, 
    and $G$ is nondecreasing, then there exists $x' > \sigma_H$ 
    such that $G(x) > H(x), \forall x \in (\sigma_H, x')$, 
    then we have 
    \begin{align*}
        \int_0^{x'} G(x)dx 
        = \int_0^{\sigma_H} G(x)dx + \int_{x'}^{\sigma_H} G(x)dx 
        & >  \int_0^{\sigma_H} H(x)dx + \int_{x'}^{\sigma_H} H(x)dx \\
        & = \int_0^{x'} H(x)dx ~,
    \end{align*}
    which violates the definition of $H \blackwellOrder G$.
    
    \item 
    Suppose that $G(\sigma_H) < H(\sigma_H)$, 
    as $H$ is continuous over $[0, 1]$,
    and $G$ is nondecreasing, then there exists $x' < \sigma_H$
    such that $H(x) > G(x), \forall x \in (x', \sigma_H)$, 
    then consider
    \begin{align*}
        \int_0^{\sigma_H} H(x)dx 
        = \int_0^{x'} H(x)dx + \int_{x'}^{\sigma_H} H(x)dx
        & > \int_0^{x'} G(x)dx + \int_{x'}^{\sigma_H} G(x)dx\\
        & = \int_0^{\sigma_H} G(x)dx~,
    \end{align*}
    which violates the condition that $\sigma_G = \sigma_H$. 
\end{itemize}
\end{proof}

\begin{proof}[Proof of \Cref{thm_r_utility_sigma}]
To prove \Cref{thm_r_utility_sigma}, we 
use the following 
result which characterizes 
the best payoff that any central planner 
can possibly hope to achieve.
Fix a \infoStrategy\ $G$ and its corresponding $\sigma_G$, 
define following {\em capped value}:
\begin{align*}
    \kappa_G := \min \{x, \sigma_G\}, ~~ x \sim G~. 
\end{align*}
Given a strategy profile $(G_1, \ldots, G_n)$,
the below lemma shows that 
the optimal agent's payoff is the highest 
capped value among senders.
\begin{lemma}
[\citet{KWE-16}]
\label{agent_opt}
The procedure defined in \Cref{thm_optimal_searching}
can achieve the agent's optimal expected payoff 
$\expect{\max_i \kappa_{G_i}}$, i.e., 
the highest expected capped value he obtains.
\end{lemma}
Recall that $\agentU(G_i, G_{-i})$ 
denote the agent's expected payoff 
when the agent is using the 
{\em optimal} inspection strategy, i.e., 
$\agentU(G_i, G_{-i}) = 
\expect[G_1, \ldots, G_n]{\max_i \kappa_{G_i}}$.

We are now ready to prove \Cref{thm_r_utility_sigma}.
We first observe that for any \infoStrategy\ $G$ 
such that $\expect[x\sim G]{x} = \lambda$, 
we have $\expect[G]{\kappa_G} = \lambda - c$.
To see this, note that
\begin{align*}
    \expect[G]{\kappa_G} 
    = \int_0^{\sigma_G} xd G(x) + \sigma_G \int_{\sigma_G}^1dG(x)
    = \lambda - c~.
\end{align*}
Given a strategy profile $(G_i, G_{-i})$, 
from \Cref{agent_opt},
the agent's optimal expected payoff is 
the expectation of the maximum of $n$ 
independent random variables $\{\kappa_{G_i}\}_{i\in[n]}$
where each random variable $\kappa_{G_i}$ has the mean $\lambda_i - c_i$. 
Let $\kappa_{G_{-i}} := \{\kappa_{G_1}, \ldots, \kappa_{G_{i-1}}, 
\kappa_{G_{i+1}}, \ldots, \kappa_{G_n}\}$.
Now observe that, 
\begin{align*}
   \expect[G_1, \ldots, G_n]{\max_i \kappa_{G_i}}
   = \expect[G_{-i}]{\expect[G_i]{\max \left\{\kappa_{G_i}, \kappa_{G_{-i}}\right\}}}~.
\end{align*}
We first prove the ``only if'' direction.
Below, we first show that for all possible $\kappa_{G_{-i}} = b$, 
the following holds
\begin{align}
    \label{stronger_goal}
    \expect[G_i']{\max \left\{\kappa_{G_i}, b\right\}} \ge 
    \expect[G_i]{\max \left\{\kappa_{G_i}, b\right\}}~.
\end{align}
Recall that from \Cref{prop_informativness_sigma}, 
we have $\sigma_{G_i'} \ge \sigma_{G_i}$. 
We now consider the following two cases:  
\begin{itemize}
\item 
When $ b \geq  \sigma_{G_{i}}$, we have 
$\expect[G_i]{\max \left\{\kappa_{G_i}, b\right\}} = b$, 
and  $ 
\expect[G_i']{\max \left\{\kappa_{G_i}, b\right\}}
\ge 
b$, 
thus \eqref{stronger_goal} holds true.

    \item 
    When $b < \sigma_{G_i}$, in this case, we have
    \begin{align*}
        \expect[G_i]{\max \left\{\kappa_{G_i}, b\right\}}
        & = \int_0^b b dG(x) + \int_b^1 \max\{\kappa_{G_i}, b\} dG_i(x)\\
        & \labelrel={thm_3_9_helper_1} bG(b) + \lambda_i - c_i - \int_0^b x dG_i(x)
        \labelrel={thm_3_9_helper_2} \lambda_i - c_i + \int_0^b G_i(x)dx~,
    \end{align*}
    where equality \eqref{thm_3_9_helper_1} uses the 
    earlier observation 
    $\expect[G_i]{\kappa_{G_i}} = \lambda_i - c_i$, 
    and equality \eqref{thm_3_9_helper_2} uses integration by parts.
    Recall that $G_i'$ is an MPS of $G_i$, we have $\int_0^b G_i(x)dx \le \int_0^b G_i'(x)dx, \forall b$.
    As a result, we conclude that $\expect[G_i]{\max \left\{\kappa_{G_i}, b\right\} } \le 
    \expect[G_i']{\max \left\{\kappa_{G_i}, b\right\}}$.
\end{itemize}
Putting all pieces together, \eqref{stronger_goal} 
holds for any $b \in [0, 1]$, 
which completes the proof for the ``only if'' direction.

We now prove the ``if'' direction. 
Recall that from \Cref{agent_opt},
$\agentU(G_i', G_{-i}) \ge \agentU(G_i, G_{-i})$ is equivalent to
$\expect[G_{-i}]{\expect[G_i']{\max \left\{\kappa_{G_i'}, \kappa_{G_{-i}}\right\}}} 
\ge \expect[G_{-i}]{\expect[G_i]{\max \left\{\kappa_{G_i}, \kappa_{G_{-i}}\right\}}}$.
Now consider a no information strategy $G_j$ for every 
box $j$ where $j\neq i$.
Then we have $\kappa_{G_j} = \lambda_j - c_j$. 
We now choose the mean $\lambda_j = \expect[x\sim G_j]{x}$
and the cost $c_j$ for each box $j$ 
such that $b \equiv \lambda_j - c_j, \forall j\in[n]\setminus\{i\}$ 
for some $b \in [0, 1]$. 
Notice that we can vary $\lambda_j, c_j, \forall j\in [n]\setminus \{i\}$
to ensure that $b$ can 
take any value between $0$ and $1$.
Then $\expect[G_{-i}]{\expect[G_i']{\max \left\{\kappa_{G_i'}, \kappa_{G_{-i}}\right\}}} 
\ge \expect[G_{-i}]{\expect[G_i]{\max \left\{\kappa_{G_i}, \kappa_{G_{-i}}\right\}}}$ for all $G_{-i}$ and all cost $(c_i)_{i\in[n]}$
implies that the following holds
\begin{align}
    \label{ineq:intermid}
    \expect[G_i']{\max \left\{\kappa_{G_i'}, b\right\}}
    \ge 
    \expect[G_i]{\max \left\{\kappa_{G_i}, b\right\}}, 
    \quad \forall b \in [0, 1]~.
\end{align}
Suppose the mean $\expect[x\sim G_i]{x} = \expect[x\sim G_i']{x} = \lambda_i$, 
consider a cost $c_i$ for box $i$ satisfying $c_i = \lambda_i$, 
then we have the reservation value for box $i$ satisfying
$\sigma_{G_i} = 0$. 
Thus, 
$\expect[x\sim G_i]{\max \left\{\kappa_{G_i}, b\right\}}
= \expect[x\sim G_i]{\max \left\{x, b\right\}}
= \lambda_i + \int_0^b G_i(x)dx$. 
Similarly, we also have 
$\expect[G_i']{\max \left\{\kappa_{G_i'}, b\right\}}
= \lambda_i + \int_0^b G_i'(x)dx$. 
Thus, From inequality \eqref{ineq:intermid}, 
we have 
\begin{align*}
    \int_0^b G_i'(x)dx \ge \int_0^b G_i(x)dx, \quad 
    \forall b \in [0, 1]~.
\end{align*}
Recall the fact that both $G_i, G_{i'}$ has the same mean $\lambda_i$,
this implies that $\int_0^1 G_i'(x)dx = \int_0^1 G_i(x)dx$.
Namely, the above inequality holds as equality for $b = 1$.
Then from \Cref{MPS_defn}, we conclude that 
distribution $G_i'$ is an MPS of distribution $G_i$.
\end{proof}

\end{document}